\documentclass[twocolumn,showpacs,preprintnumbers,amsmath,amssymb]{revtex4}

\usepackage{psfrag}
\usepackage{graphicx}
\usepackage{dcolumn}
\usepackage{bm}
\usepackage{setspace}
\usepackage{rotating}

\newcommand{\citetwo}[2]{\cite{#1},\cite{#2}}

\newcommand{\mx}{\mbox{\bf x}}

\newcommand{\muu}{\mbox{\bf u}}

\usepackage{color}
\usepackage{epsfig}

\begin{document}

\preprint{APS/123-QED}

\title{A sufficient condition for Gaussian departure in turbulence 
}

\author{Daniela Tordella$\natural$}
\email{daniela.tordella@polito.it}
\author{Michele Iovieno$\natural$}
\author{Peter Roger Bailey$\sharp$}
\affiliation{
$\natural$ Dipartimento di Ingegneria Aeronautica e Spaziale, Politecnico di Torino, Corso Duca degli Abruzzi 24, 10129 Torino, Italy}%
\affiliation{$\sharp$   Scuola di Dottorato, Politecnico di Torino, Corso Duca degli Abruzzi 24, 10129 Torino, Italy}%


\date{\today}

\begin{abstract}



The interaction of two isotropic turbulent fields of equal integral scale but different kinetic energy generates the simplest kind of inhomogeneous turbulent field. In this paper we present a numerical experiment where two time decaying isotropic fields of kinetic energies $E_1$ and $E_2$  initially match over a narrow region. 
Within this region the 
kinetic energy varies as a hyperbolic tangent. The following temporal evolution produces a shearless  mixing.  The  anisotropy and intermittency of velocity and velocity derivative statistics is observed. In particular the asymptotic behavior in time and as a function of the energy ratio $E_1/E_2 \rightarrow \infty$ is discussed. This limit  corresponds to the maximum observable turbulent energy gradient for a given $E_1$ and is obtained through the limit $E_2 \rightarrow 0$. A field with $E_1/E_2 \rightarrow \infty$ represents a mixing which could be observed near a surface subject to a very small velocity gradient separating two turbulent fields, one of which is nearly quiescent. In this condition the turbulent penetration is maximum and reaches a value equal to 1.2 times the nominal mixing layer width.
The experiment  shows that the presence of a turbulent energy gradient is  sufficient  for the appearance of intermittency and  that during the mixing process the pressure transport is not negligible  with respect to the turbulent velocity transport. These findings may open the way to the hypothesis that the presence of a gradient of turbulent energy is the minimal requirement for Gaussian departure in turbulence.

\end{abstract}
\maketitle 

\section{\label{introduction}Introduction}
\begin{figure}
\vspace*{2mm}
\psfrag{D}{\Large$2\Delta$}
\psfrag{E1}{\Large$E_1$}
\psfrag{E2}{\Large$E_2$}
\includegraphics[width=0.8\columnwidth]{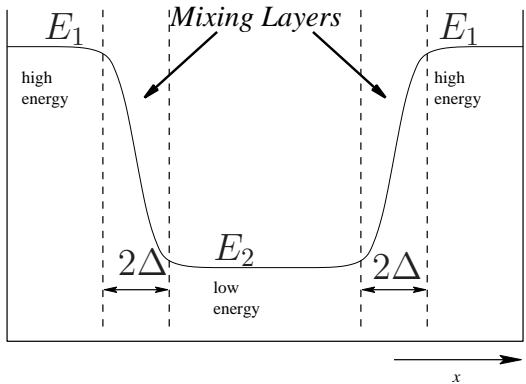}
\caption{
Scheme of the flow. Direction $x$ is the mixing direction. The high energy ($E_1$) and low energy ($E_2$) regions are separated by mixing layers of conventional thickness $\Delta(t)$ defined   by mapping the low energy side of the mixing layer to zero and the high energy side to one. $\Delta(t)$ is  equal to the distance between the points with normalized energy values 0.25 and 0.75 \cite{vw89}, \cite{ti06}.}
\label{fig.schema1}
\end{figure}

\begin{figure}
\psfrag{u}{$v$}
\includegraphics[width=0.8\columnwidth]{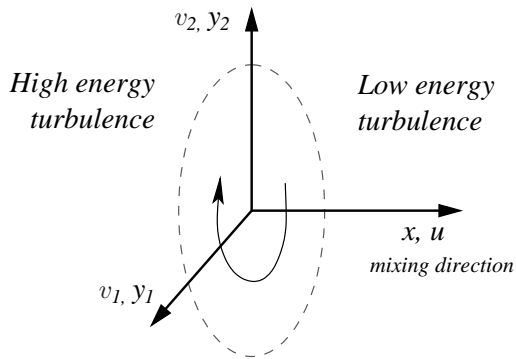}
\caption{Scheme of the flow. Reference frame:  $y_1,y_2$ are normal to  $x$, that is the direction of the flow inhomogeneity. The flow is homogenous in all planes normal to this direction.
}
\label{fig.schema2}
\end{figure}

\begin{figure*}
\begin{minipage}[h]{0.83\textwidth}
\includegraphics[width=0.51\columnwidth,angle=-90]{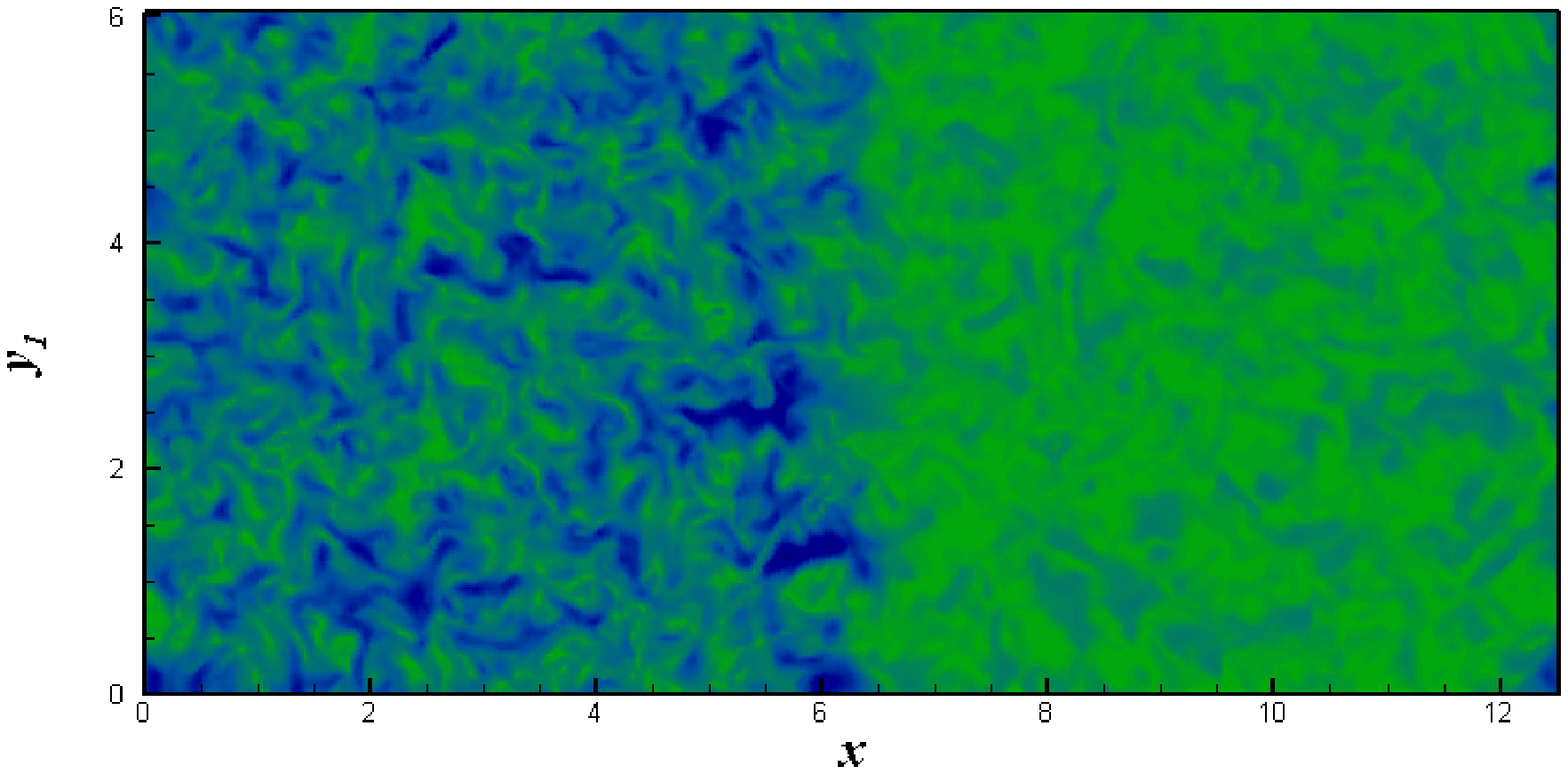}
\includegraphics[width=0.51\columnwidth,angle=-90]{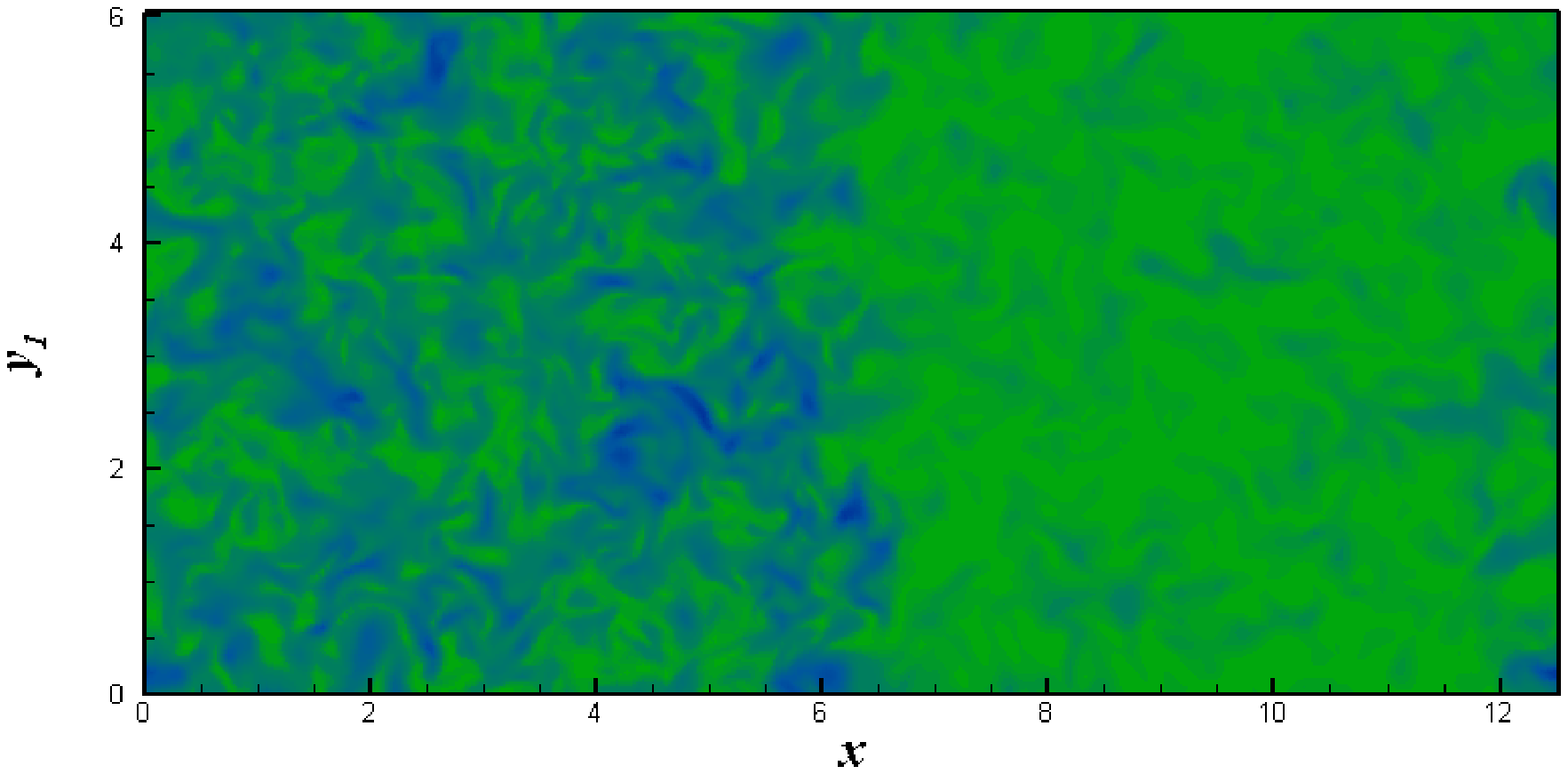}
\end{minipage}
\hfill
\begin{minipage}[h]{0.15\textwidth}
\includegraphics[width=0.45\columnwidth,angle=0]{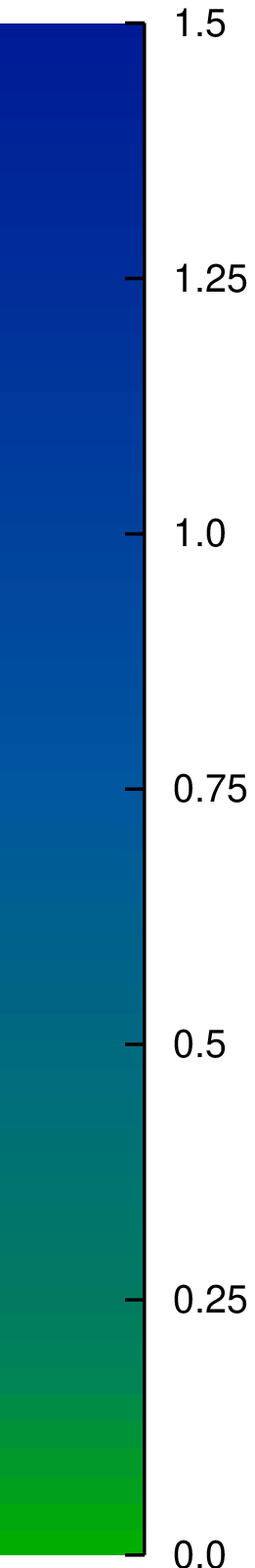}
\end{minipage}
\caption{(Color online) Visualization at two time instants of contours of kinetic energy $E(x,y_1,y_2,t)/E_1(0)$ in a plane at constant $y_2$, $E_1/E_2=6.7$, $Re_\lambda=45$: (a) $t/\tau=0.8$, (b) $t/\tau=2.5$.}
\label{fig.livelli}
\end{figure*}

A turbulent shearless mixing layer is generated by the interaction of two homogeneous isotropic turbulent (HIT) fields, see definition diagrams in figures \ref{fig.schema1} and \ref{fig.schema2} and the flow visualizations  in figure \ref{fig.livelli}. This kind of mixing is characterized by the absence of a mean shear, so that there is no production of turbulent kinetic energy and no mean convective transport. The turbulence spreading is caused only by the fluctuating pressure and velocity fields. The  inhomogeneous statistics  are typically due to the presence of the gradients of turbulent kinetic energy and integral scale. The shearless turbulence mixing was first experimentally investigated by Gilbert (1980)\cite{g} and by Veeravalli and Warhaft (1989) \cite{vw89} by means of passive grid generated turbulence. Later on, numerical investigations were carried out by Briggs {\sl et al.} (1996) \cite{bfkm96} and Knaepen {\sl et al.} (2004) \cite{kdc}, and more recently by Tordella and Iovieno (2006, 2007) \cite{ti06,it07}. All these studies considered a decaying turbulent mixing.

In all studies, apart from that of Gilbert, where the turbulent energy ratio was very low, the mixing layer was observed to be highly intermittent and the transverse velocity fluctuations seen to have large skewness. Across the mixing the distributions of the second, third and fourth order moments collapse when the mixing layer width is used as lengthscale \cite{vw89,ti06,it07}. 

In passive grid laboratory experiments the gradients of integral scale and kinetic energy are intrinsically linked. In past studies the ratio of the integral scale of the interacting turbulence fields was in the range 1.3 \cite{g} - 4.3 \cite{vw89} with a ratio of kinetic energies in the range 1.5 \cite{g} - 23 \cite{vw89}. In numerical \cite{ti06} or active grid experiments these two parameters can be independently varied.

In the present study, a mixing configuration in which the integral scale is homogeneous is considered. The ratio of the turbulent kinetic energies has been chosen as the sole control parameter and is varied from $1.5$ to $10^6$, $Re_\lambda$ of the high turbulent energy field is 45. The aim of this study is to show the  intermittent behavior of such a configuration that in the past was considered to have almost Gaussian velocity statistics. This interpretation was motivated by the absence of both a kinetic energy production and an integral scale variation, two typical sources of intermittency and was also supported by laboratory observations carried out in the absence of  a sufficiently high kinetic energy gradient \cite{g}. 
Another aim of this numerical experiment is to reach the asymptotic condition where the kinetic energy ratio ${\cal E}= E_1/E_2$ goes to infinity. 
This last condition is relevant in applications concerning the diffusion of a turbulent field in a region of quiescent fluid, where extreme bursts of rate of strain and vorticity can be expected \cite{lm05}. The presence of such events is shown  by high values of skewness and kurtosis.

A  description of the numerical experiment is given in section II. Data on the degree of anisotropy observed in the second and third order velocity moments are described in section III, where an interpretation based on Yoshizawa's hypothesis is also given. In section IV we present the two types of asymptotics  considered: the temporal asymptotics of the second and third order velocity moments, and the asymptotics with respect to the turbulent kinetic energy ratio of the velocity skewness, mixing penetration and kurtosis. In addition,  a smaller set of data on the temporal asymptotics of third and fourth order moments of the velocity derivative is also discussed in this last section. The concluding remarks are presented in section V.

\section{\label{setup}Numerical experiment}

\begin{figure}
\includegraphics[width=0.8\columnwidth]{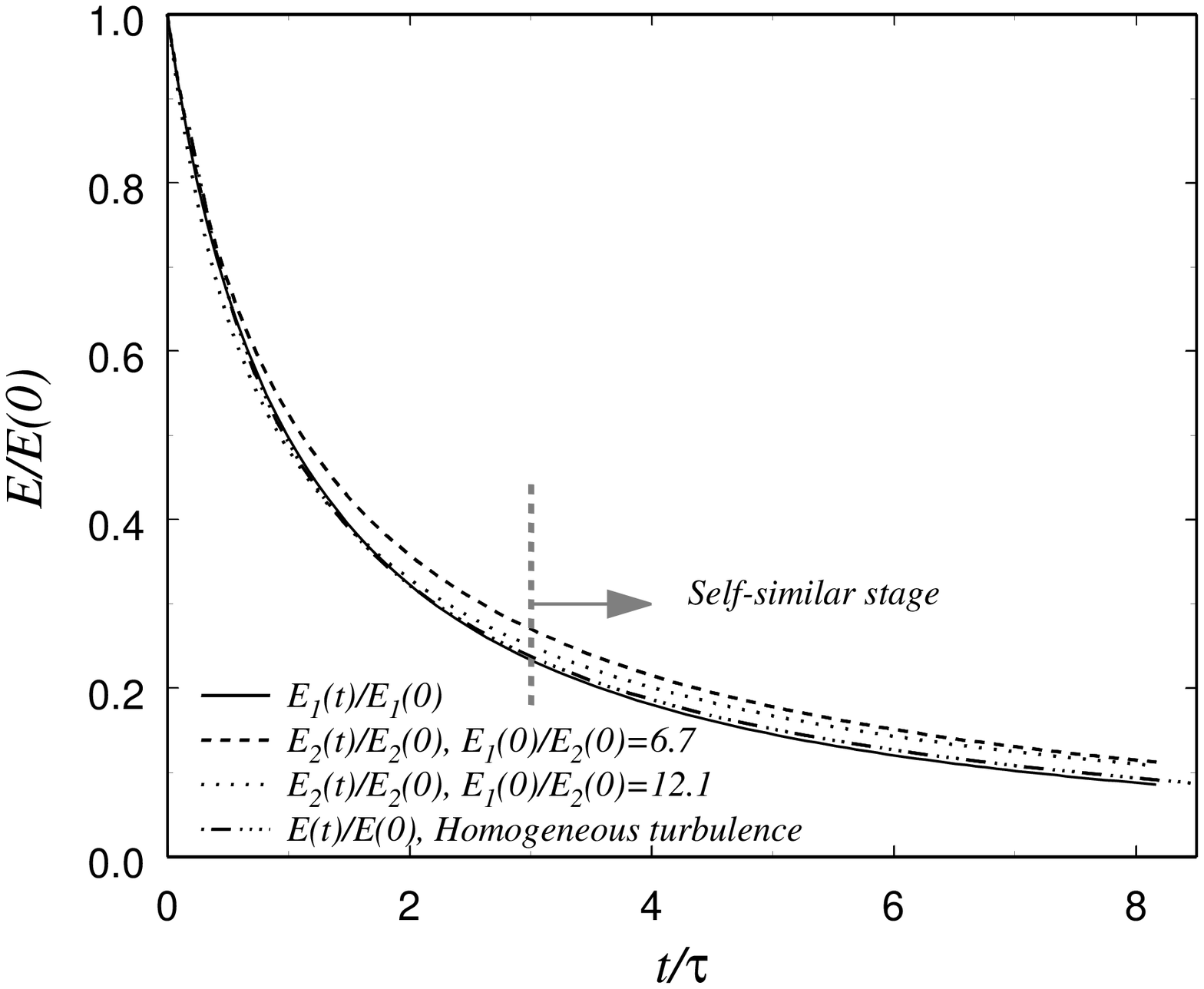}
\vskip -5.45cm \hskip +2.77cm
\includegraphics[width=0.35\columnwidth]{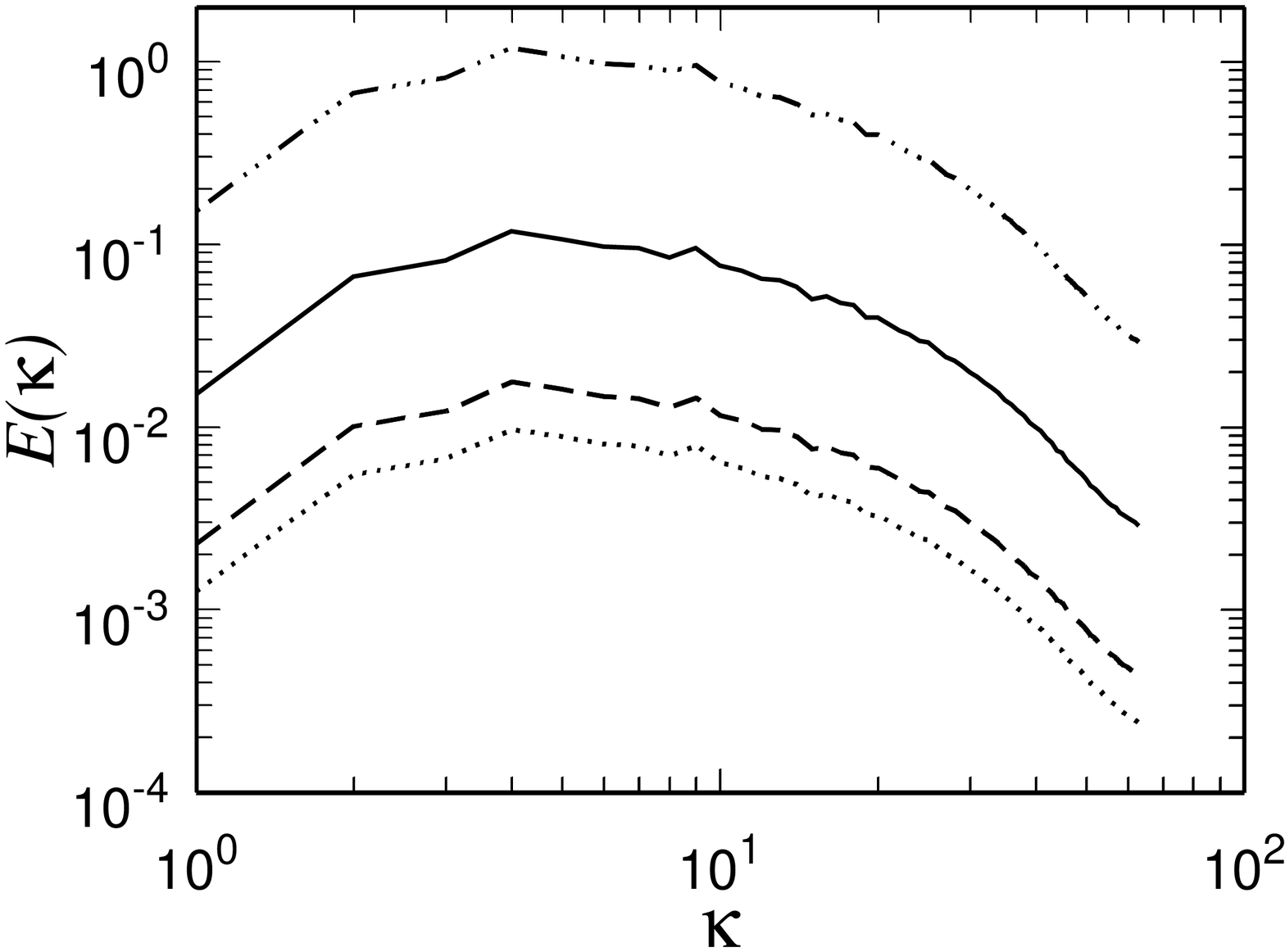}
\vskip 2.6 cm
\caption{Turbulent kinetic energy decay of the two interacting isotropic turbulent fields ($E_1$ high energy, $E_2$ low energy) at $re_\lambda=45$ and, in the inset, the corresponding initial energy spectrum. Data from a homogeneous and isotropic turbulence ($E$) simulated in a $(2\pi)^2\times8\pi$ domain ($128^2\times 512$) have been shown for comparison. The initial spectrum is equal to the spectrum of the high energy region in the mixing.}
\label{fig.decadimento}
\end{figure}



Navier-Stokes equations are numerically solved with a fully dealiased (3/2 - rule) Fou\-rier-Ga\-lerkin pseudospectral method \cite{ict01}. 
The computational domain is a parallelepiped with periodic boundary conditions in all directions, see fig.1.  Tests were performed on a  $4\pi(2\pi)^2$ parallelepiped domain with $256\times128^2$ points. Further tests with a $8\pi(2\pi)^2$ parallelepiped with $512\times128^2$ points were used to obtain an estimate of the numerical accuracy.  The Taylor-microscale Reynolds number $Re_\lambda$, corresponding to the high energy field, is equal to 45 for  both the spatial discretization of the direct numerical simulations (DNS). 

In the initial condition, the two isotropic turbulent  fields  are matched by means of a hyperbolic tangent function. This transition layer represents 1/40 of the $4\pi$ domain, and 1/80 of the $8\pi$ domain. The matched field is 

\begin{equation}
\muu(\mx)=\muu_{1}({\mx})p(x) + \muu_{2}({\mx})(1-p(x))
\label{u0}
\end{equation}

\begin{eqnarray}
p(x)&=&\frac{1}{2}\left[1+\tanh\left(a\frac{x}{L} \right)\tanh\left(a 
\frac{x-L/2}{L}\right)\times\right.\nonumber\\ &\times&\left.  \tanh\left( a\frac{x-L}{L} \right)\right]
\end{eqnarray}


\noindent where the suffixes $1,2$ indicate high and low energy sides of the mixing respectively, $x$ is the inhomogeneous direction, $L$ is the width of the computational domain in the $x$ direction.
Constant $a$ in (\ref{peso}) determines the initial mixing layer thickness $\Delta$, 
conventionally defined as the distance between the points with normalized energy values 0.25 and 0.75 when the low energy side is mapped to zero and the high energy side to one. When $ a = 12 \pi $ the ratio $\Delta/L$ is about $0.026$, for $ a = 20\pi$ the ratio $\Delta/L$ is about $0.015$.
These values have been chosen so that this initial thickness is large enough to be resolved but small enough to have large regions of homogeneous turbulence during the simulations. 
This technique of generating  the transition layer is analogous to that used in Briggs et al. (1996)\cite{bfkm96}, and Knaepen et al. (2004)\cite{kdc}.
The matching on which the initial condition is built up is a linear superposition of the two isotropic fields as indicated in equations (\ref{u0}) and (\ref{peso}). 
A set of statical preperties of the high  kinetic energy HIT field is shown in Table 1. Since the low energy field  $u_2$ is  obtained by multiplying
the  initial velocity field $u_1$ by a constant,  the numerical experiment carried out by
mixing these fields is  a turbulent mixing with different energies but of equal integral scale. It should be noted that, by doing so, the mean pressure along the mixing direction is not constant, However, the mean pressure gradient is opposite to the gradient of turbulent kinetic energy and thus no mean velocity field is generated, see the appendix. Examples of the shearless mixing obtained in this way for direct numerical simulation can be found  in \cite{bfkm96} and \cite{ti06}. The initial spectra of the two HIT fields are shown in figure \ref{fig.decadimento}.
In this figure the temporal decay of the two isotropic turbulent fields is shown together with, as a reference, the decay of the homogeous and isotropic turbulence simulated in one of the computational domain used to simulate  the turbulent shearless mixing ($(2\pi)^2\times8\pi$, $128^2\times512$). In figure \ref{fig.decadimento} the estimate of the time instant where the self-similar decay of the mixing starts is also shown.


Let us now consider the flow symmetry. It can be seen that a shearless mixing is a flow in which only one direction of inhomogeneity is present, as a consequence any plane normal to the inhomogeneous direction is homogeneous. This corresponds to a cylindrical symmetry. See the reference frame scheme in figure \ref{fig.schema2}.

The time integration is carried out by means of a four-stage fourth-order explicit Runge-Kutta scheme.
Statistics are obtained by averaging over planes normal to the inhomogeneous direction, see figure \ref{fig.schema2}.

The initial conditions were generated from the homogeneous and isotropic turbulent field produced by Wray in 1998 \cite{wray}, which is a classic data set often used in literature.

A posteriori, it is possible to obtain numerical accuracy estimates. 
The raw data by Wray has an inhomogeneity level on the kinetic energy of
about $\pm 8\%$ 
and skewness and kurtosis values slightly different from those of the statistical equilibrium ($0.02 \pm 0.12$ instead of 0 and $2.8 \pm 0.2$ instead of 3, respectively). As far as our set of direct numerical simulations is concerned, the increase in width of the computational domain from $4\pi$ to $8\pi$ (from 256 to 512 grid points) allowed  an estimate of the relative accuracy to be obtained. For the maximum values of the distributions across the mixing, the accuracy is of about 5\% for the skewness, and of about 8\% for the  the kurtosis. 
%

In figure \ref{fig.skewness-kurtosis}, which summarizes the results regarding the  maximum values reached by the velocity skewness and kurtosis within the mixing  and the results about the penetration, it can be seen that the simulations with initial $\Delta/L=1/40$ and $1/80$ yield data which collapse in a satisfactory way. 
On checking the symmetry of the numerical solutions, which, due to the periodicity of the boundary conditions, contain two mixings, see scheme in figure \ref{fig.schema1}, it was verified that the doubling of the computational domain induces a decrease of the asymmetry  from 10\% to 5\% for the skewness and from 20\% to 15\% for the kurtosis.

\begin{table}
%
\begin{tabular}{rll}
\multicolumn{3}{c}{Velocity statistics}\\[2pt]
\hline\hline\\[-1mm]
\multicolumn{1}{c}{$E_1$}   &  \multicolumn{1}{c}{$S_1$}   & \multicolumn{1}{c}{$K_1$} \\[2pt] \hline\\[-1mm]
~$1.01\pm0.08$ & ~$1.6\cdot10^{-2}\pm0.12$  & ~$2.85\pm0.2$\\
\hline\hline
\end{tabular}\\[12pt]
\begin{tabular}{llll}
\multicolumn{4}{c}{Velocity derivative statistics}\\[2pt]
\hline\hline\\[-1mm]
  \multicolumn{1}{c}{$S_{\partial u/\partial x}$} & \multicolumn{1}{c}{$K_{\partial u/\partial x}$}
& \multicolumn{1}{c}{$S_{\partial u/\partial y,1}$} & \multicolumn{1}{c}{$K_{\partial u/\partial y,1}$}\\[2pt] \hline\\[-1mm]
  ~$-0.42 \pm0.08$ & ~$3.61\pm0.2$ 
& ~$-0.40 \pm0.08$ & ~$3.53\pm0.2$ \\
\hline\hline
\end{tabular}
\caption{Statistical properties of the high energy HIT field $u_1$ at $t=0$. $E_1$ is the normalized turbulent kinetic energy, $S_1$ and $K_1$ are the velocity skewness and kurtosis, $S_{\partial u/\partial x}$ and   $K_{\partial u/\partial x}$ are the velocity longitudinal derivative skewness and kurtosis,
$S_{\partial u/\partial y,1}$ and   $K_{\partial u/\partial y,1}$ are the velocity transversal derivative skewness and kurtosis.
The field $u_1$ was obtained from the data base by Wray 1998 \cite{wray}. Note, that these statistical properties are the same for all the considered low energy fields $u_2$, since they were obtained by multiplying the initial high energy field by a constant. }
\end{table}

\section{\label{Anisotropy}Anisotropy and Yoshizawa's hypothesis}

\begin{figure}
\includegraphics[width=0.7\columnwidth]{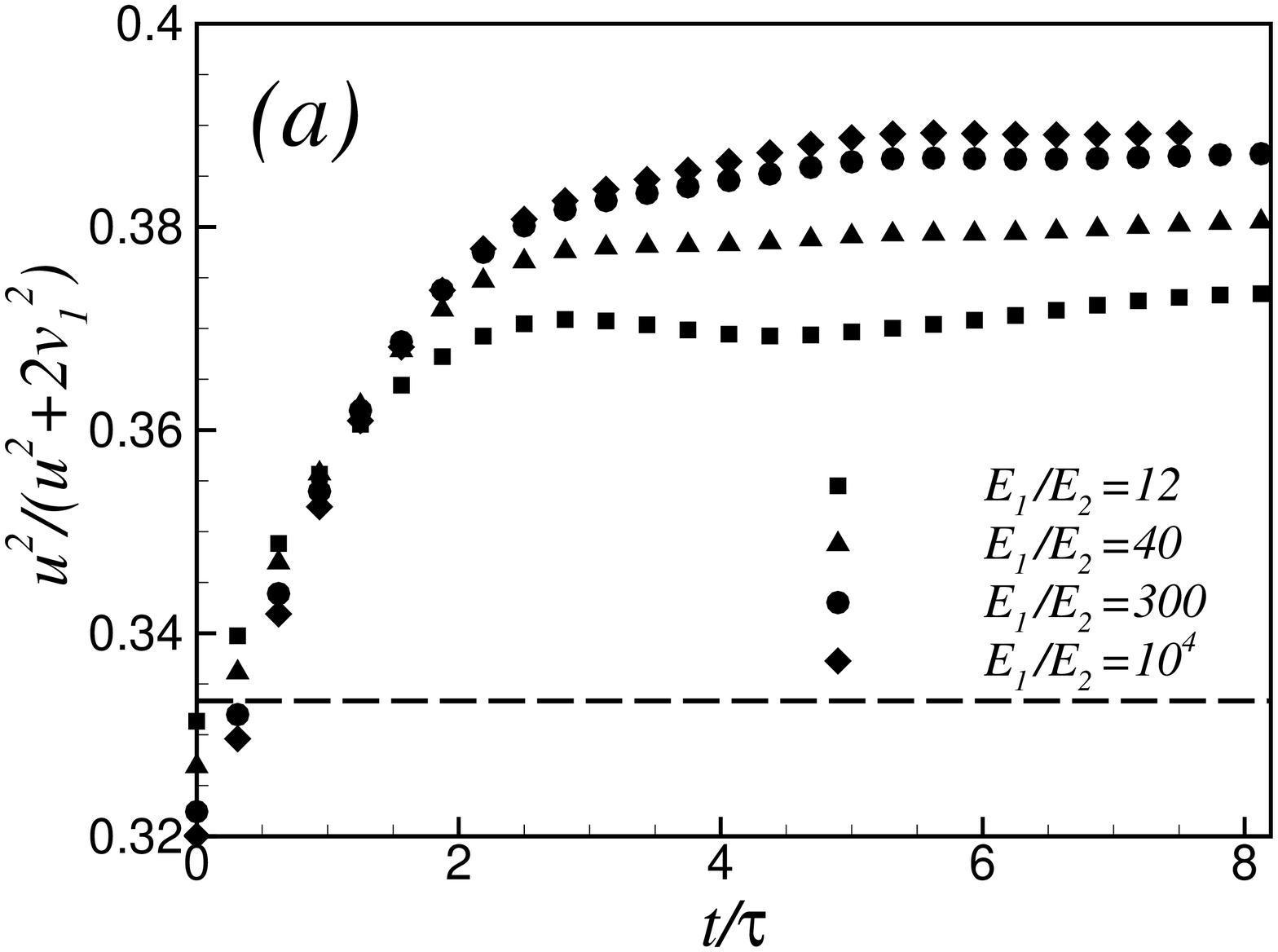}\\
\vspace*{-4cm}
\hspace*{-0.68\columnwidth}
\rotatebox{90}{{\colorbox{white}{~~$\overline{u^2}/\overline{\left(u^2+v_1^2+v_2^2\right)}$~~}}}\\
\vspace*{0.6cm}
\includegraphics[width=0.7\columnwidth]{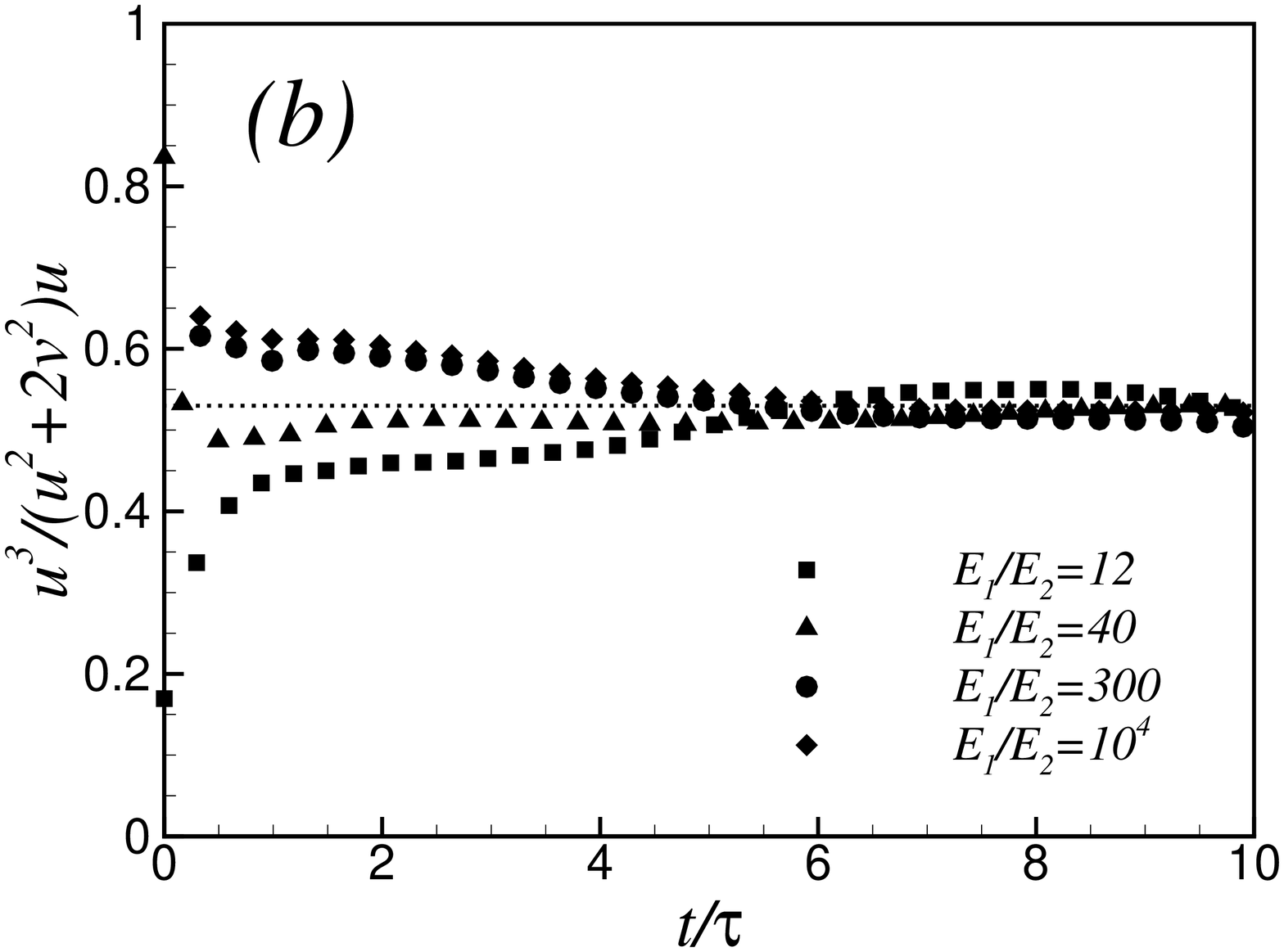}\\
\vspace*{-4cm}
\hspace*{-0.67\columnwidth}
\rotatebox{90}{{\colorbox{white}{~~$\overline{u^3}/\overline{(u^2+v_1^2+v_2^2)u}$~~}}}
\vspace*{0.1cm}

\caption{Anisotropy of the turbulent second and third order moments at the centre of the mixing layer. The horizontal dashed line in part (a) indicates the isotropic reference value, the horizontal dotted line in part (b) indicates the estimate of the asymptote value.}
\label{fig.isotropia}
\end{figure}

%

In isotropic turbulence the normalized second order moment of the velocity components, normalized with the sum $\overline{u^2}+ \overline{v_1^2 } + \overline{v_2^2}$,  is $1/3$, whilst the third order moment is zero. In the present flow the field anisotropy develops during the mixing process. The value of the normalized moments vary in time and reach an asymptotic value after few time units, see figure \ref{fig.isotropia}. The time unit  $\tau$ is defined as $\tau = \ell(0)/E_1^{1/2}(0)$), where $\ell$ is the integral scale, here uniform across the mixing,  and $E_1$ is the turbulent kinetic energy of the high energy side of the mixing.

An  initial turbulent energy gradient  $\nabla E = (E_1 - E_2)/ (2 \Delta)$ corresponds to each value of ${\cal E}= E_1/E_2$.  The width $\Delta$  is defined  by mapping the low energy side of the mixing layer to zero and the high energy side to one, and it is equal to the distance between the points with energy values 0.25 and 0.75, as in the paper by Veeravalli \& Warhaft \cite{vw89} (in the following referred to as V\&W).  The turbulent energy gradients can be normalized  by  the  value of the high energy field, and by the value of the mixing thickness $\Delta(t)$. It should be noticed that by doing so, the normalized gradient value has the upper limit of $0.5$, which is reached in the limit for $E_2$ going to zero.

In figure   \ref{fig.isotropia}(a) the time evolution inside the mixing of the second order moment $\overline{u^2}/(\overline{u^2}+ \overline{v_1^2 } + \overline{v_2^2})$ is shown. After a linear growth the curves bend toward the asymptotic value, which is in the range 0.37-0.39 for a kinetic energy ratio growing from 4  to $10^4$ (this corresponds to a normalized gradient of turbulent kinetic energy from $0.37$ to $0.50$, or, by supposing a mixing in air with a $Re_\lambda=45$ in the high energy side, 
to a dimensional gradient from $1.8$ to $2.4$ m/s$^2$).
%
%
%

As a consequence of the cylindrical symmetry of this mixing, it follows that the second order moments $\overline{v_1^2}/(\overline{u^2}+ \overline{v_1^2 } + \overline{v_2^2})$, $\overline{v_2^2}/(\overline{u^2}+ \overline{v_1^2 } + \overline{v_2^2})$ are equal  and  range from 0.315 to 0.305 when ${\cal E}$ varies from $4$ to $10^4$. The anisotropy level, defined as the difference between the second-moment values referred to the isotropic value,  can be considered  mild (16 \% for $ {\cal E}=12$, 25 \% for  ${\cal E}=10^4$) given that the accuracy in the original data base used to build the initial condition \cite{wray} is  of about  $8\%$ as far as both the homogeneity and isotropy are concerned. It should be considered that this level of initial accuracy of homogeneity and isotropy is excellent in nominal HIT numerical fields. In higher resolution fields $(1024^3)$  the accuracy is analogous \cite{toschi}.

Figure \ref{fig.isotropia}(a) indicates that the value 0.39 for $\overline{u^2}/(\overline{u^2}+ \overline{v_1^2 } + \overline{v_2^2})$ is reached by increasing  ${\cal E}$ from 12 to $10^4$. This value can be considered as an approximation of the asymptotic value attainable by increasing the turbulent energy gradient. 

It is important to note that in literature concerning the shearless mixing,  almost all authors report a near homogeneity in the second-order velocity moments regardless the observation method  used, numerical or laboratory \cite{vw89,bfkm96,ti06}. 

The anisotropy of the third-order velocity moments is more enhanced than that of the second-moments. This can be observed in figure \ref{fig.isotropia}(b), where the time evolution of the third order moment $\overline{u^3}$ normalized with the total kinetic energy flow in the mixing direction,  $\overline{u^3}+ \overline{v_1^2 u} + \overline{v_2^2 u}$, is plotted.
The  estimate of the temporal  asymptotic value  we obtained    is $0.53 \pm 0.03$ and does not depend on ${\cal E}$. If the level of anisotropy is defined as the difference between the third moments divided by their mean, an anisotropy of 80\% is obtained. This means that, for all the energy ratios, nearly one half of the turbulent kinetic energy flow across the mixing is due to the self transport of $\overline{u^2}$.   Let us note that at the initial instant, when the mixing process starts, the quantity $\overline{u^3}/(\overline{u^3}+ \overline{v_1^2 u} + \overline{v_2^2 u})$ is not defined because both the numerator and the denominator are not defined. This is numerically verified through the large dispersion of the initial values associated to different ${\cal E}$. Of course, this dispersion is also due to the non perfect homogeneity of the HIT data base used to build the initial condition, see section II. The  data dispersion is however reduced as the mixing process advances. After 6 times scales is less than $10\%$. 
 
It is possible to analyze this result by means of  simplifying hypotheses currently found in literature - (a) the pressure transport is almost proportional to the convective transport associated to the fluctuations (Lumley 1978\cite{tl72}, Yoshizawa, 1982, 2002, \citetwo{y82}{y02}), - (b) the dissipative scales are nearly isotropic \cite{my75}, and - (c)  the second order moments are almost isotropic as observed in shearless turbulent mixings and also confirmed by the present numerical experiment, as discussed above.


Let us now consider the one point second order moment equations

\begin{eqnarray}
\partial_t \overline{u^2} + \partial_x \overline{u^3} &=& - 2\rho^{-1} \partial_x \overline{pu} + 2 \rho^{-1}\overline{p\partial_x u} - 2\varepsilon_u + \nu\partial^2_x \overline{u^2}
\label{uu}\\
\partial_t \overline{v_i^2} + \partial_x \overline{v_i^2u} &=&  2 \rho^{-1}\overline{p\partial_{y_i}v_i} - 2\varepsilon_{v_i} + \nu\partial^2_x \overline{v_i^2}, \; i=1,2 \label{vi}
\end{eqnarray}

\noindent where $u$ is the fluctuating velocity in the inhomogeneous direction $x$,  $ v_1, v_2$ are the fluctuation components in the plane normal to $x$ and $\varepsilon_u, \varepsilon_{v_i} $ are the dissipation terms in the mixing and normal directions, respectively.

The pressure strain terms $\overline{p\partial_{x}u}$ and $\overline{p\partial_{y_i}v_i}$, in the absence of a mean flow, are of the order
of $\frac{\displaystyle{\varepsilon}}{b}(\rho\overline{u_i u_j} - \frac{2}{3}\rho b \delta_{ij})$, see for instance Monin \& Yaglom, 1971\cite{my71} (Volume 1, equation 6.12, page 379), where ${\varepsilon}$ is the total dissipation and $b$ is the turbulent kinetic energy per unit of mass.
Since, as previously explained, experiments  show no appreciable difference in the second order moments in the mixing, see condition (c) above, 
the pressure strain terms are neglected.\\ \indent
Condition (a) implies that we can write

\begin{equation}
-\overline{pu} = \alpha \rho\frac{\overline{u^3} + 2\overline{v_i^2u}}{2}\;
\label{y}
\end{equation}
\noindent  for any value of position $x$ along the mixing and for any time instant $t$. 
The difference between equation (\ref{uu}) and equation (\ref{vi}) gives
\begin{equation}
\partial_t (\overline{u^2}-\overline{v_i^2} )  +
\partial_x (\overline{u^3}-\overline{v_i^2u} ) \approx
-2\rho^{-1}\partial_x \overline{pu}
- 2(\varepsilon_{u}-\varepsilon_{v_i}).
\label{pu-derivata0}
\end{equation}
By condition (b) $\overline{u^2}\approx\overline{v_i^2}$ and by condition (c) $\varepsilon_{u}\approx\varepsilon_{v_i}$. Thus, the unsteady term on the left hand side as well as the second term on the right hand side can be neglected and it follows that
\begin{equation}
\partial_x (\overline{u^3}-\overline{v_i^2u} ) \simeq -2\rho^{-1}\partial_x \overline{pu}.
\label{pu-derivata}
\end{equation}
Integration of (\ref{pu-derivata}) with respect to $x$ leads to
$$\overline{u^3}-\overline{v_i^2u}\simeq -2\rho^{-1}\overline{pu} + C,$$
but, considering that all quantities in this equation vanish outside the mixing (i.e. for $x\rightarrow\pm\infty$), the integration constant $C$ is equal to zero. Thus 
\begin{equation}
\overline{u^3}-\overline{v_i^2u}\simeq -2\rho^{-1}\overline{pu},
\label{pu-mom-terzi}
\end{equation}
By inserting the previous  relation into (\ref{y}), it is  possible to write
%
\begin{equation}
\overline{v_i^2u} = \beta \,\overline{u^3}, \;\;\; \beta = \frac{1-\alpha}{1+2\alpha}.
\label{v2u}
\end{equation}
Then, by defining $\Phi$ the proportion of the turbulent kinetic energy flow associated to the $u$ fluctuation, it follows that
\begin{equation}
\Phi = \frac{\overline{u^3}}{(\overline{u^3}+ 2\overline{v_i^2 u})} = \frac{1}{1+2 \beta}
\label{anis-3}.
\end{equation}

We have computed the constant $\alpha$ for  the present experiments and found, in asymptotic temporal condition and for ${\cal E} \in [12, 10^4]$, an average value of $0.37 \pm 0.03$. This gives $\beta \sim 0.36$ and $\Phi \sim 0.58$ 
This last value contrasts with our numerical experimental value of $\Phi=0.53 \pm 0.03$ shown in fig.5 b.

We have verified that  $\alpha$ and $\beta$ remain almost constant during the decay and when varying the shearless mixing parameter ${\cal E}$, a fact which confirms that the pressure transport correlation is almost proportional to the convective transport associated to the fluctuations and confirm the Yoshizawa hypothesis that when the turbulent field does not posses a unidirectional mean flow, the velocity turbulent transport term is not dominating the pressure transport \cite{y82, y02, lmk97}. In the present mixing both  the advection and the production rate of the turbulent energy are zero and thus the turbulent transport (velocity and pressure) rate is of the same order of the dissipation rate. 


\begin{figure}
\includegraphics[width=0.85\columnwidth]{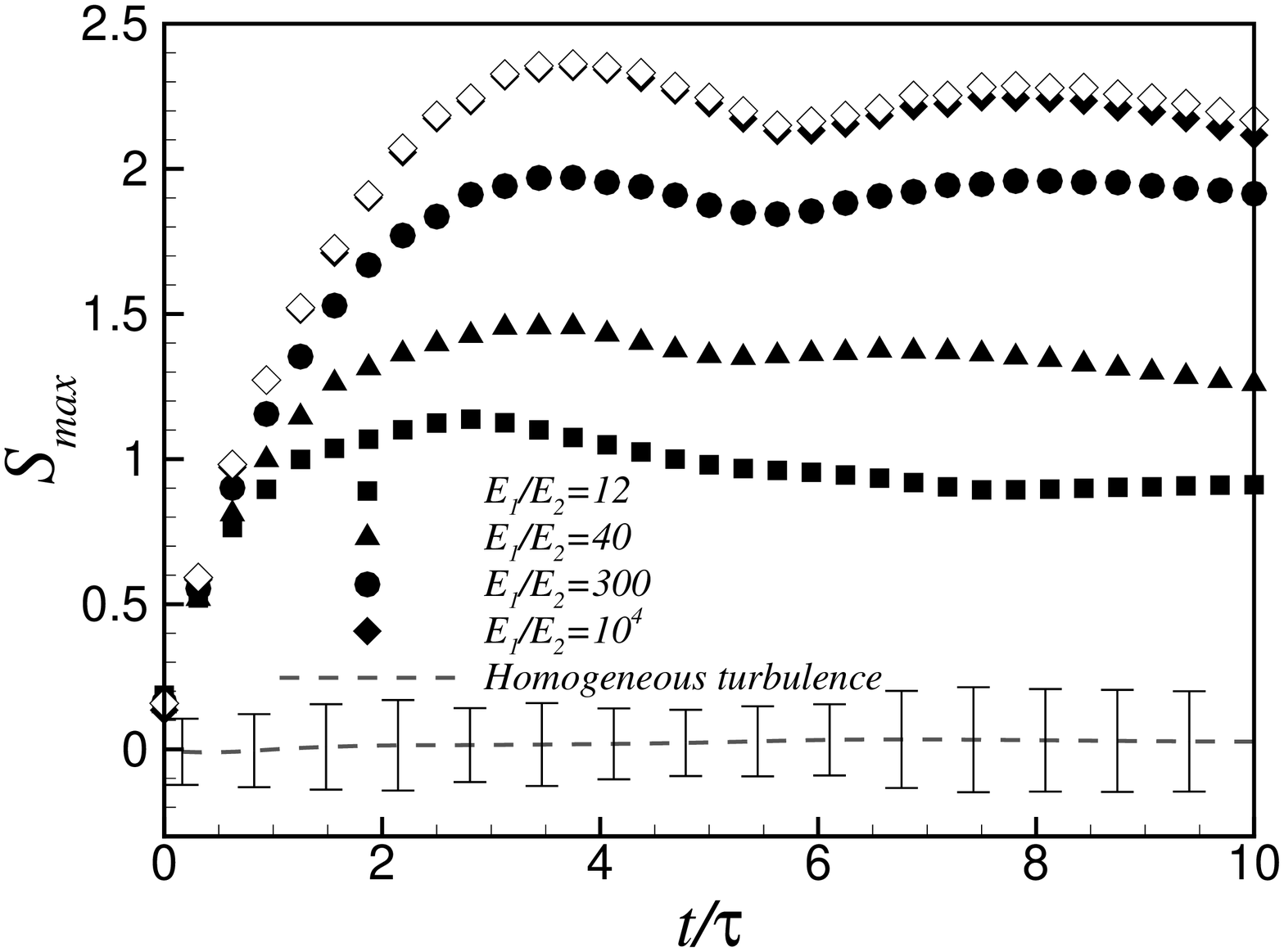}\\ 
\includegraphics[width=0.85\columnwidth]{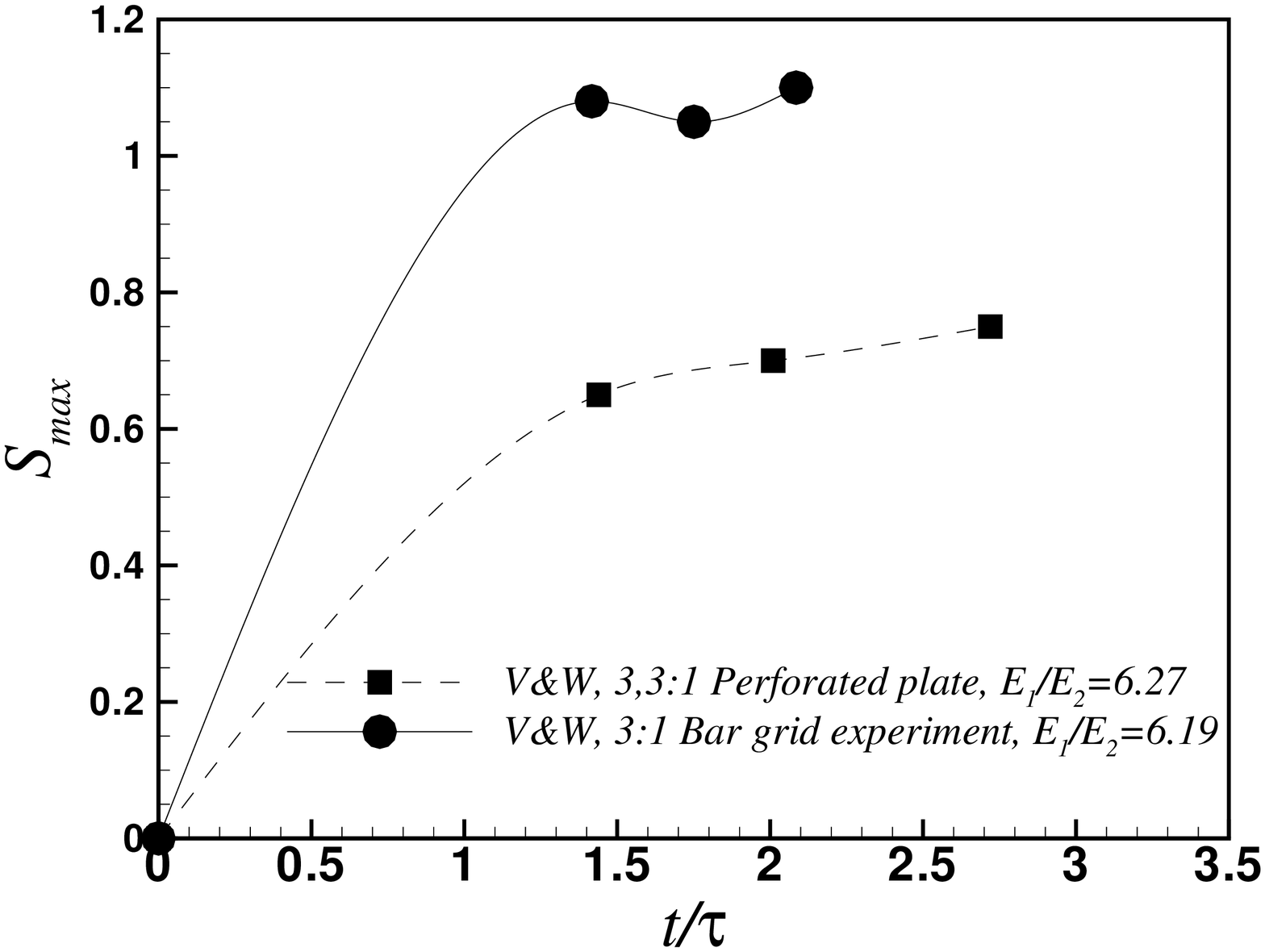}
\caption{Temporal evolution of the maximum of the skewness $S=\overline{u^3}/\left(\overline{u^3}\right)^{3/2}$ in the mixing for various energy ratios ranging from $12$ to $10^4$.
(a) Numerical experiments at $Re_\lambda=45$. Empty symbols refer to simulations in a $(2\pi)^2\times8\pi$ domain, the others to simulations in a $(2\pi)^2\times4\pi$ domain. The dashed line is the value of the reference skewness in a simulation of homogeneous and isotropic turbulence carried out on the same computational domain, bars represent the maximum fluctuations of this skewness. (b) Laboratory data at $Re_\lambda=44.5$ (perforated plate experiment) and  $Re_\lambda=78.1$ (bar grid experiment) from wind tunnel experiments where a spatial decay is observed \cite{vw89}. The time in laboratory experiments has therefore been computed using Taylor's hypothesis, as $t$ = $d/U$, where $d$ is the distance from the grid and $U$ is the mean velocity across the grids.}
\label{fig.stempo}
\end{figure}

\begin{figure}
\includegraphics[width=0.85\columnwidth]{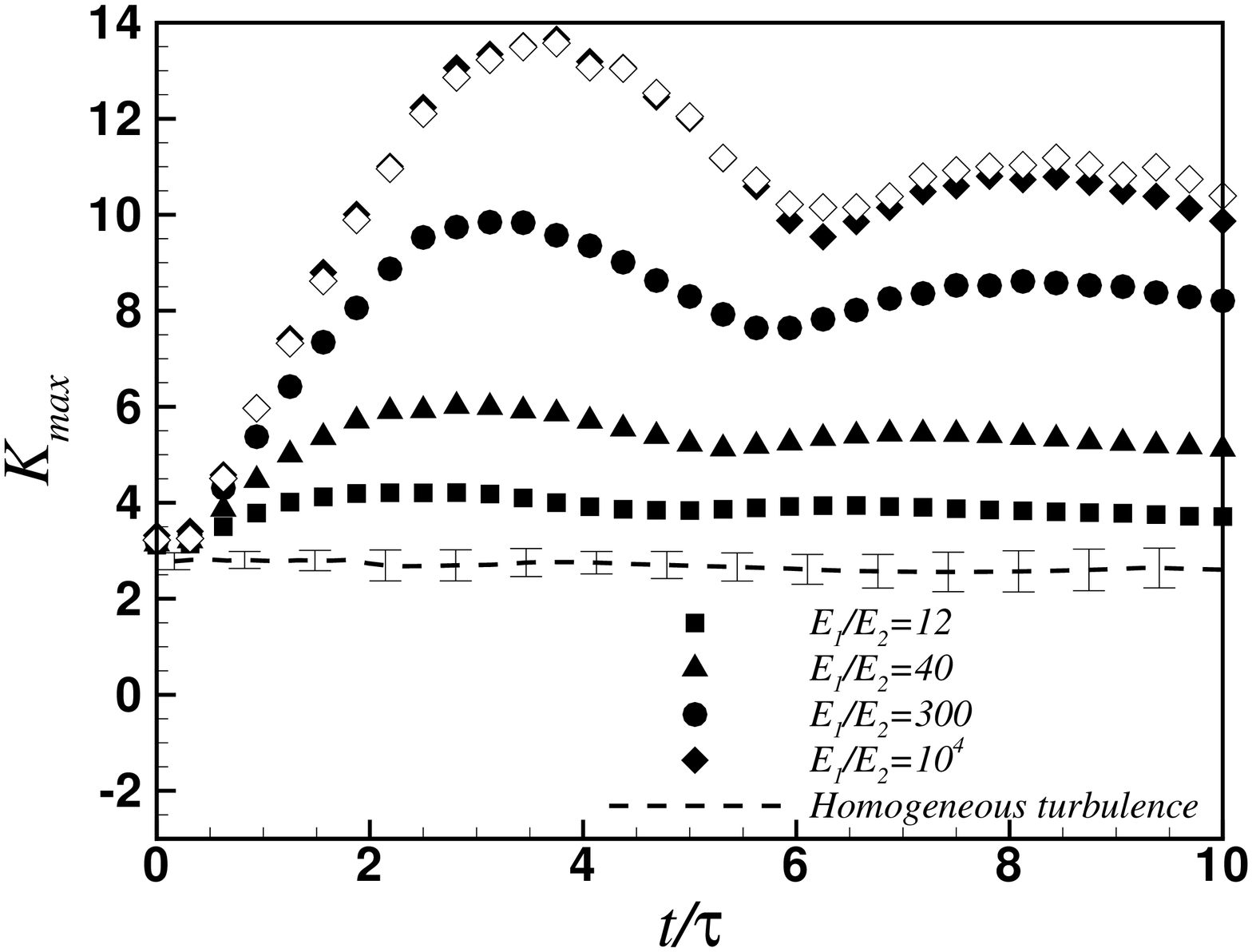}\\ 
\includegraphics[width=0.85\columnwidth]{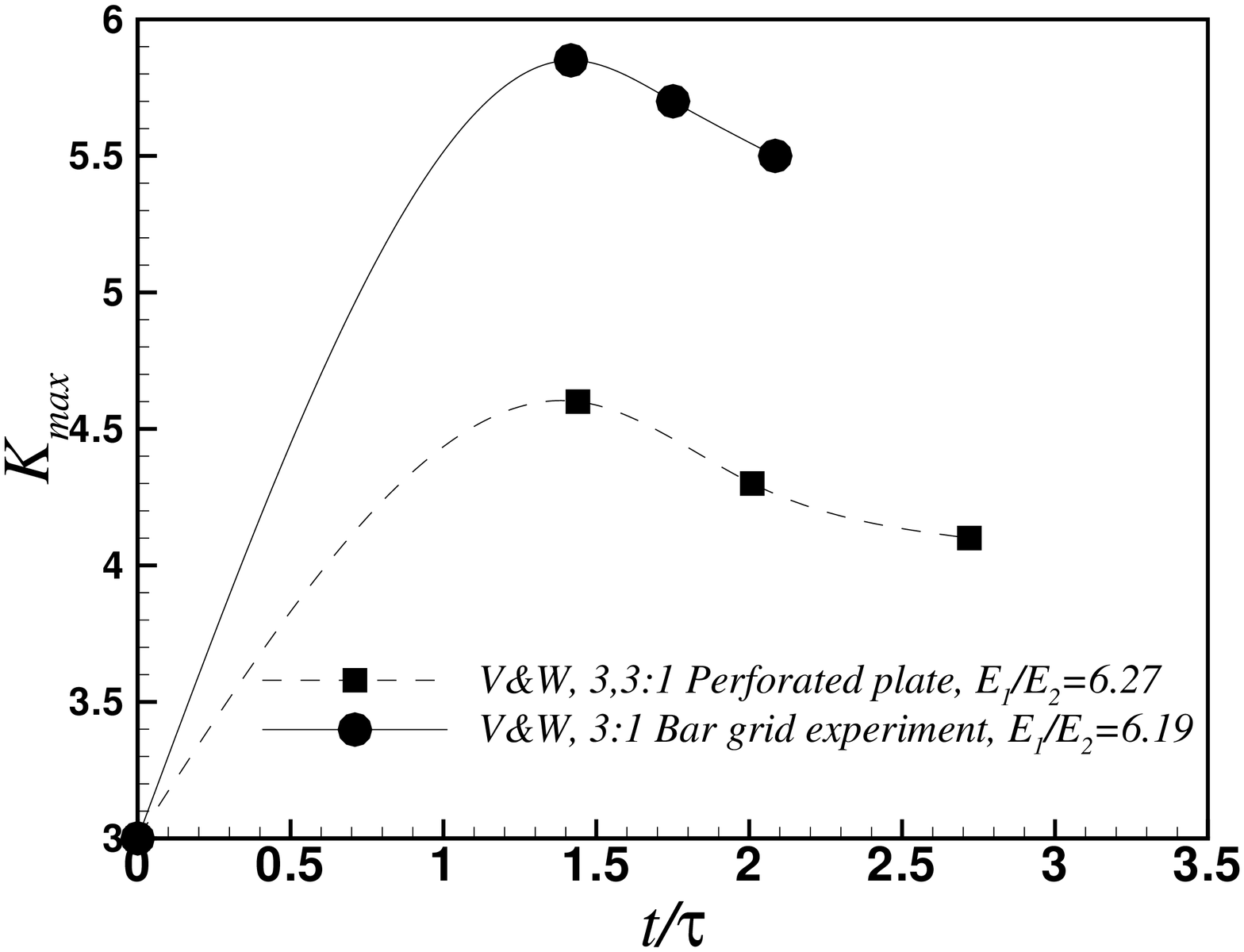}
\caption{Temporal evolution of the maximum of the kurtosis $K=\overline{u^4}/\left(\overline{u^2}\right)^{2}$ in the mixing for various energy ratios ranging from $12$ to $10^4$.
(a) Numerical experiments at $Re_\lambda=45$,
Empty symbols refer to simulations in a $(2\pi)^2\times8\pi$ domain, the others to simulations in a $(2\pi)^2\times4\pi$ domain.
The dashed line is the value of the reference kurtosis in a simulation of homogeneous and isotropic turbulence carried out on the same computational domain, bars represent the maximum fluctuations of this kurtosis. (b) Laboratory data at $Re_\lambda=44.5$ (perforated plate experiment) and  $Re_\lambda=78.1$ (bar grid experiment) from wind tunnel experiments where a spatial decay is observed \cite{vw89}. The time in laboratory experiments has therefore been computed using Taylor's hypothesis, as $t$ = $d/U$, where $d$ is the distance from the grid and
$U$ is the mean velocity across the grids.}
\label{fig.ktempo}
\end{figure}

\begin{figure}
\includegraphics[width=0.85\columnwidth]{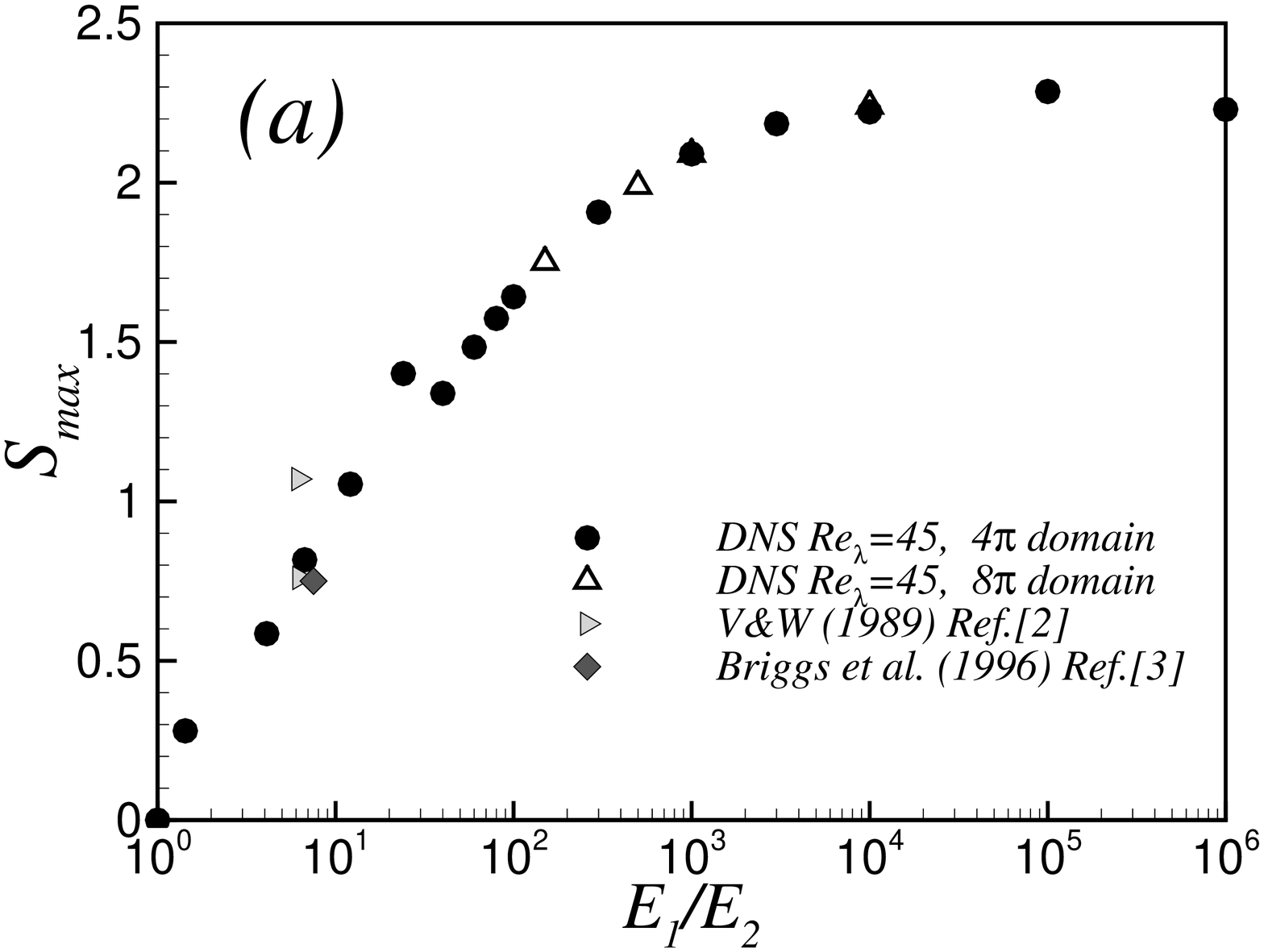}
\includegraphics[width=0.85\columnwidth]{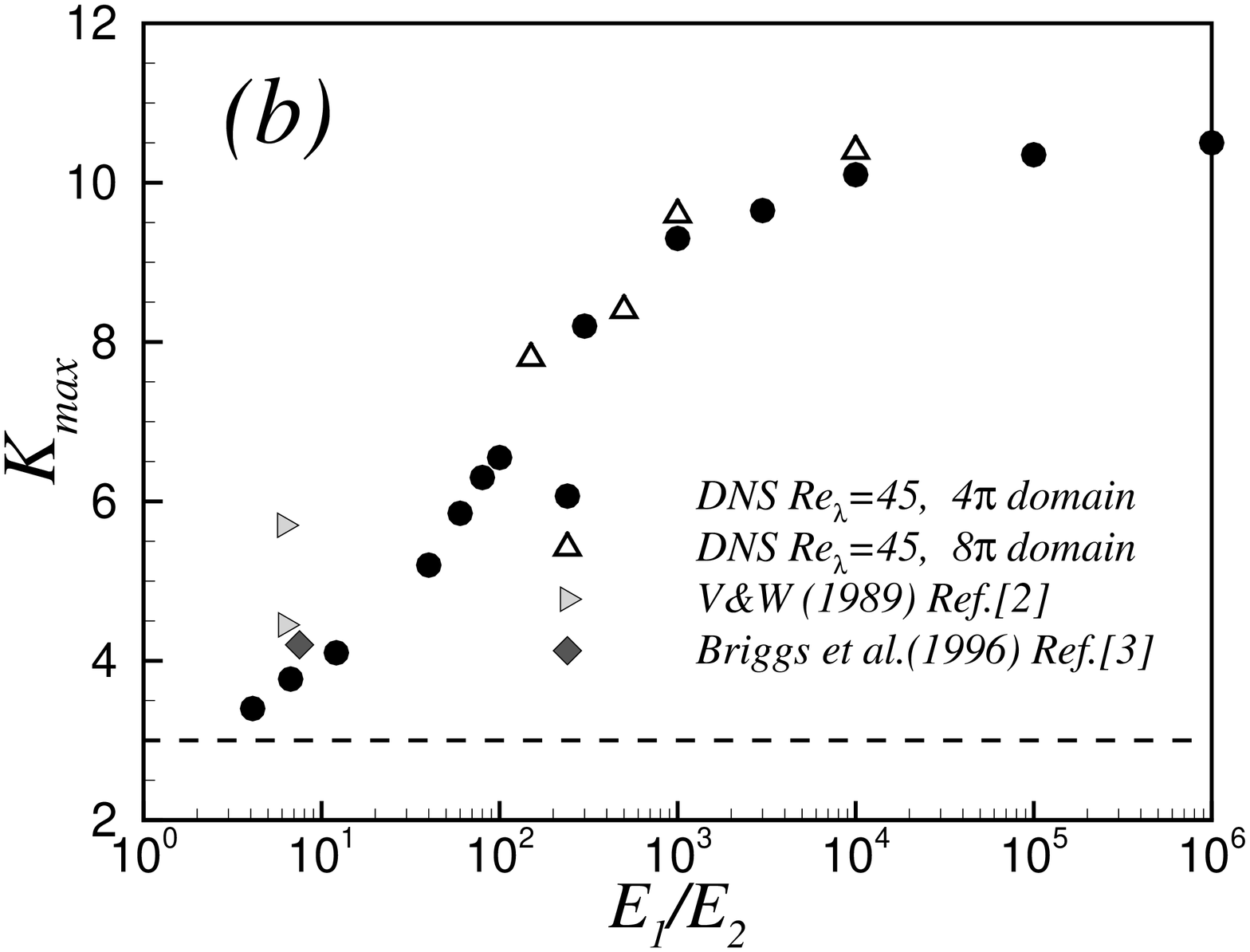}
\includegraphics[width=0.85\columnwidth]{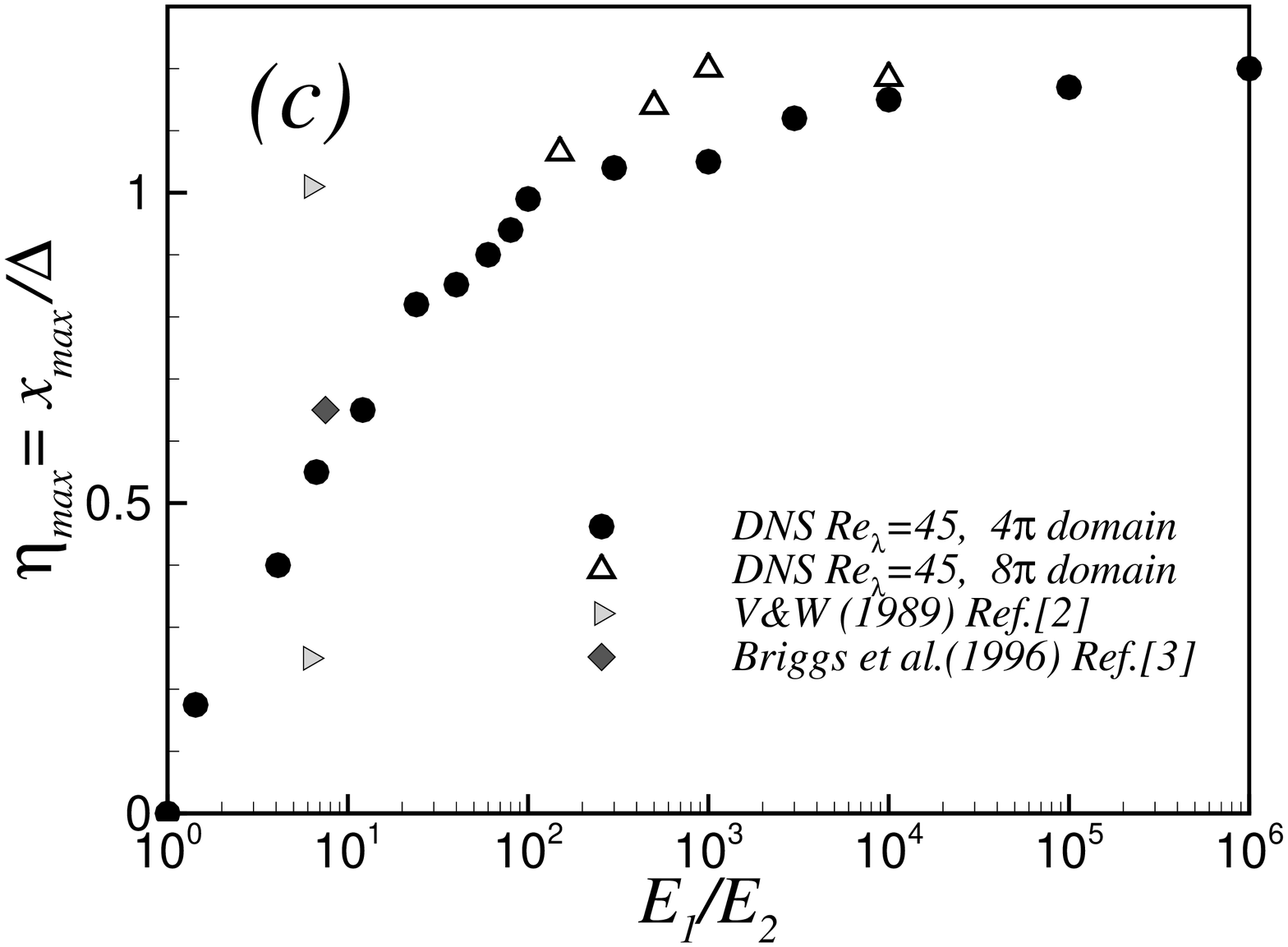}
\caption{(a) Maximum of the skewness and (b) estimate of the asymptotic kurtosis value as a function of the initial energy ratio (the horizontal dashed line indicates the Gaussian reference value), (c) normalized  position of the maximum of the skewness in the mixing layer as a function of the initial energy ratio. Note that one can expect a higher level of intermittency in the data of \cite{vw89} since these experiments had non-unity integral scale ratio ($\ell_1/\ell_2\simeq1.5$).}
\label{fig.skewness-kurtosis}
\end{figure}

\section{\label{Asymptotics}Intermittency asymptotic behavior}

In this section we consider the asymptotic behavior with regards to the variation of the parameter that controls this kind of shearless mixing layer, that is
the initial energy ratio ${\cal E}=E_1/E_2$ between the high energy turbulent field 1 and the low energy turbulent field 2. As stated  above, this ratio is
unequivocally linked to the turbulent kinetic energy gradient. In this work, ${\cal E}$ was varied between $1.5$ and $10^6$. The two external fields show, for moderate values of ${\cal E}$, decay exponents which are very close, so that the two homogeneous turbulences external to the mixing decay in a similar way and the value of $E_1/E_2$ remains quite constant during the time interval considered \cite{ti06, it07}.

After few initial eddy turnover times $\tau=\ell(0)/E_1^{1/2}(0)$, where $\ell$ is the initial integral scale (homogeneous through the whole domain) and $E_1(0)$ is the initial energy of the high energy side, a true mixing layer begins to emerge from the initial conditions and reaches a self-similar state.
This means that all normalized moments distributions across the mixing collapse to a single curve when the position is normalized with the mixing layer thickness, which is defined as the distance between the points with normalized energy $(E-E_2)/(E_1-E_2)$ equal to $1/4$ and $3/4$, see sketch in fig.1.
This definition has been used in many previous works on shearless mixing \cite{vw89, bfkm96, ti06}.

Results from numerical simulations show that the mixing layer is highly intermittent in the self-similar stage of decay, and its intermittency is dependant on ${\cal E}$. In order to analyze the flow intermittency, moments of the component $u$, that is the component in the direction of the flow of turbulent kinetic energy, were computed (the averages are computed by integrating over planes at $x = const$). A particular focus was placed on the skewness $S=\overline{u^3}/\left(\overline{u^2}\right)^{3/2}$ and kurtosis $K=\overline{u^4}/\left(\overline{u^2}\right)^{2}$. 

The velocity fluctuation $u$ is responsible for the energy transport across the mixing.
The skewness distribution is a principal indicator of intermittent behavior. It vanishes in
homogeneous isotropic turbulent flows and thus it remains close to zero in the fields external to the mixing. The skewness takes a positive value within the mixing layer. 
Figure \ref{fig.stempo}(a) shows the time evolution of the maximum of the skewness for four simulations
with energy ratios between 12 and $10^4$. 
During the initial eddy turn-over times the skewness increases
steadily,
before bending at a time varying from $1.5$  (${\cal E}= 12 $) to $4$  (${\cal E}=10^4$). 
At this point the mixing layer enters a near self-similar stage
of evolution.
Figure \ref{fig.stempo}(b) shows the time evolution of the maximum of the skewness in the  V\&W experiments,  the 3,3:1 perforated plate experiment, where ${\cal E}=6.27$, and the 3 : 1 bar grid experiment, where ${\cal E}=6.19$. Since in the laboratory all the statistics decay in space, we have estimated an equivalent temporal decay by using   Taylor's hypothesis. The corresponding time in laboratory experiments has been computed as $t$ = $d/U$, where $d$ is the distance from the grid and $U$ is the mean velocity across the grids \cite{vw89}. By comparing parts (a) and (b) of figure \ref{fig.stempo} one can see a good agreement. The distribution with the lowest value of 
${\cal E}$ in part (a), which is 12, start to bend at 1.5 - 1.7 eddy turn-over times and has values of $S_{max}$ approaching those of \cite{vw89}. Note that in the laboratory experiment the ratio of macroscales is about $1.5$ (this value is estimated by considering the finiteness of $Re_\lambda$ according to Sreenivasan (1998)\cite{s98}). This agrees with the finding \cite{ti06, it07} that if the gradient of kinetic energy and macroscale are concurrent the mixing process is enhanced. In fact, one sees here that an higher energy gradient, ${\cal E}=12$, produces the same skewness  than the gradient of scale associated with the  lower energy gradient,  ${\cal E}=6.19$, in the V\&W experiment.
In our numerical experiment, for the higher ${\cal E}$ ratios,  we note a sort of damped oscillation that appears beyond the first maximum. This seems also to be shown by the 3 : 1 bar grid experiment, see figure \ref{fig.stempo}(b). 


The value of maximum skewness inside the mixing layer as a function of the energy ratio is depicted in figure \ref{fig.skewness-kurtosis}(a). For values
of $E_1/E_2$ lower than $10^2$ it scales almost linearly with the logarithm of the energy ratio, which is in fair agreement with the scaling exponent of $0.29$ found in \cite{ti06}.


Figure \ref{fig.ktempo} shows the temporal evolution  of the maximum of the kurtosis inside the layer. Here again the comparison between our numerical data and the data of the V\&W experiment  is presented. The numerical and the laboratory results contrast well for comparable values of $E_1/E_2$. A high peak is  shown  at the end of the formation time interval where the mixing process develops. This peak is followed by a decrease, that could be interpreted as the fact that the more extreme intermittent turbulent events take place at the end of the  formation interval and before  the self-similarity sets in. In the numerical experiments which last more time scale units than those in the laboratory, the decrease is followed by another  damped increase-decrease cycle, as in the skewness case. The time asymptotic values were estimated by averaging over the last cycle. Note that data in figures \ref{fig.stempo} and \ref{fig.ktempo}   
 from laboratory experiments were obtained in the presence of  concurrent  gradients  of integral scale and kinetic energy. Also in the kurtosis case, it can be observed that a higher energy gradient produces the same intermittency than a gradient of scale associated with a lower energy gradient \cite{ti06, it07}.

The distribution of the peak of kurtosis inside the mixing  is shown in figure \ref{fig.skewness-kurtosis}(b). From this  figure it can be noted that the kurtosis reaches very high values, much higher than the value of 3, that is the Gaussian reference value indicated in the figure by the dashed line. The kurtosis asymptote is in fact close to 10.5, which indicates the presence in the mixing layer of extremely intense  intermittent events.

A similar behavior of the skewness and kurtosis maxima can be seen in the mixing penetration, defined as the instantaneous position along the $x$ direction of the maximum of the skewness normalized with the instantaneous mixing layer thickness $\Delta(t)$, see figure \ref{fig.skewness-kurtosis}(c). The penetration becomes  constant in the self-similar evolution. The penetration physically highlights the region of maximum intermittency, which is located in the low energy side of the mixing layer. An increase of the energy ratio enhances the penetration of the high energy side into the low energy side. An asymptotic value of about $1.2 \Delta$ is obtained for ${\cal E}\rightarrow\infty$, which gives an indication of the penetration of an isotropic turbulent field into a quiescent field.

\begin{figure}
\psfrag{W}{{\itshape\large (a)}}
\psfrag{Y}{$\!\!S_{\partial u/\partial x}$}
\includegraphics[width=0.49\columnwidth]{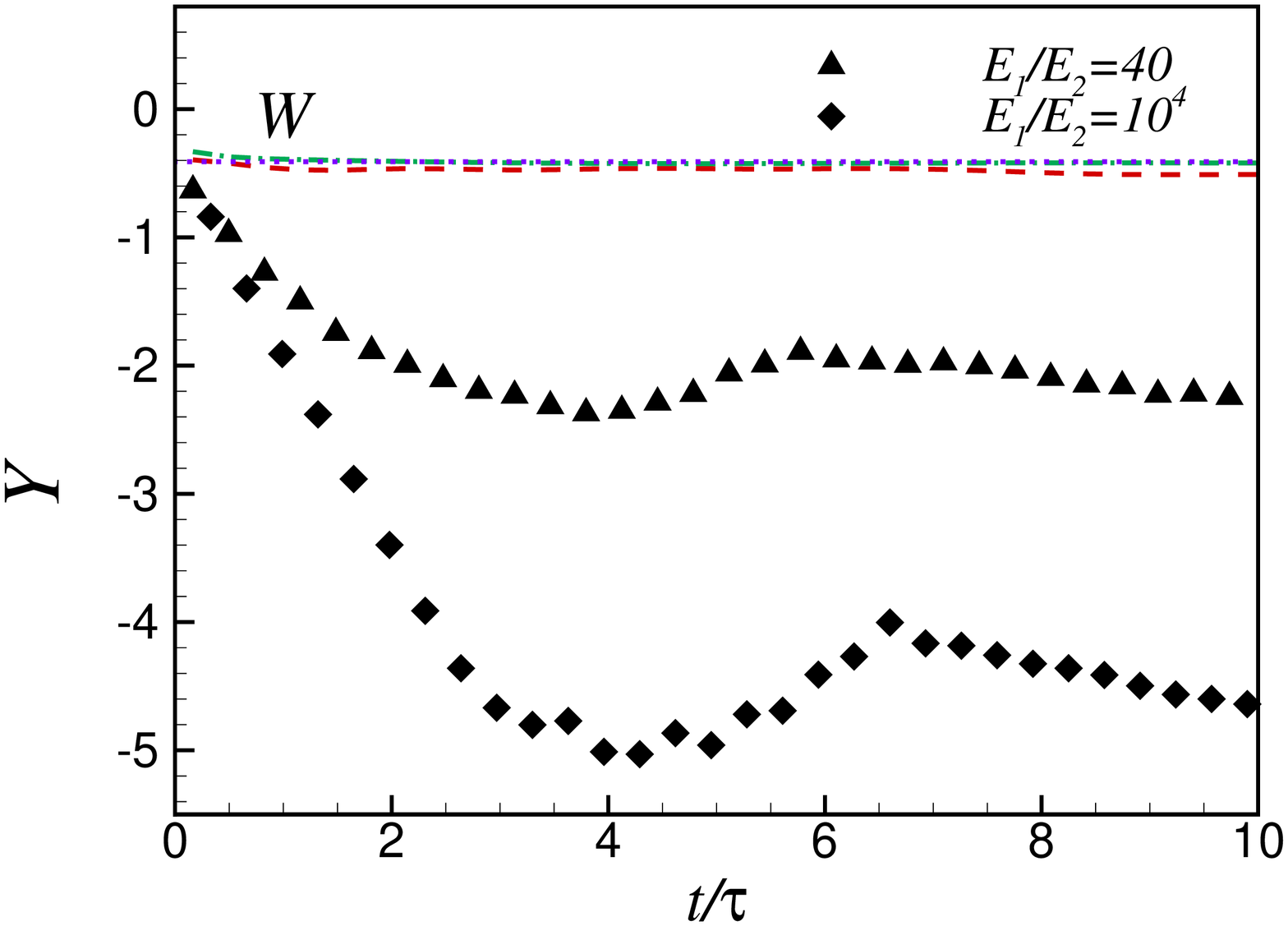}
\psfrag{W}{{\itshape\large (b)}}
\psfrag{Y}{$\!\!K_{\partial u/\partial x}$}
\includegraphics[width=0.49\columnwidth]{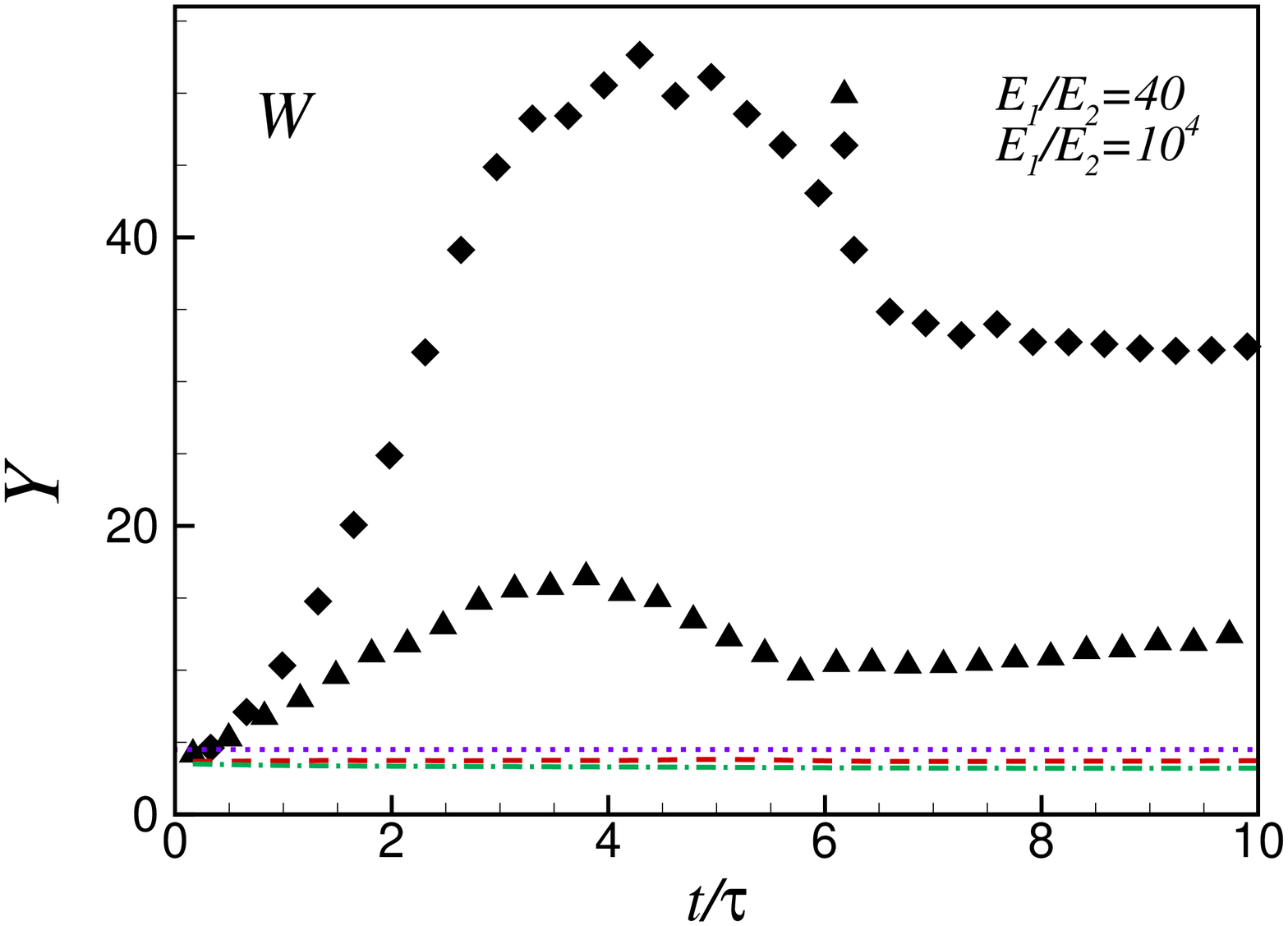}
\psfrag{W}{{\itshape\large (c)}}
\psfrag{Y}{$\!\!S_{\partial u/\partial y_{i}}$}
\includegraphics[width=0.49\columnwidth]{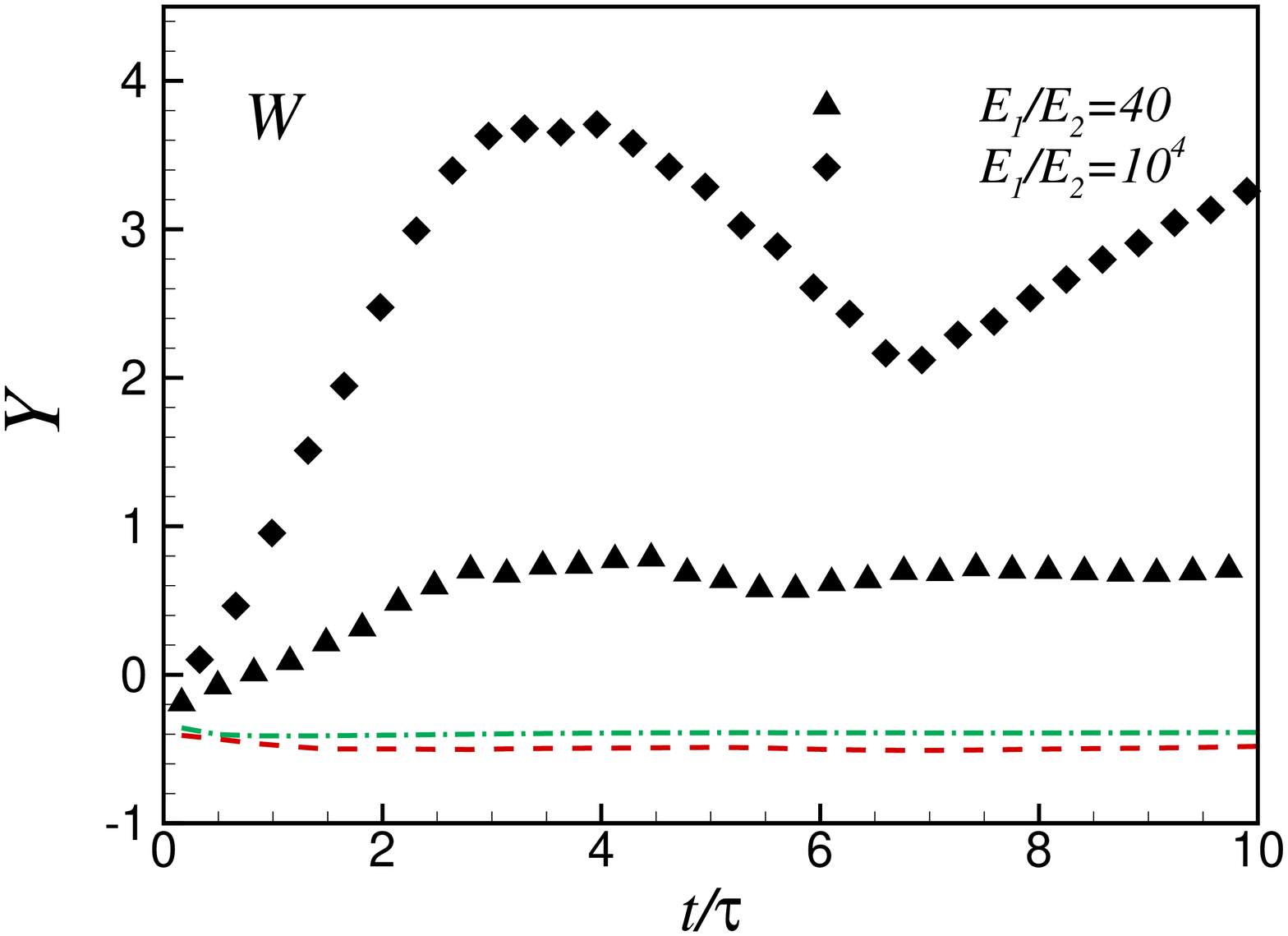}
\psfrag{W}{{\itshape\large (d)}}
\psfrag{Y}{$\!\!K_{\partial u/\partial y_i}$}
\includegraphics[width=0.49\columnwidth]{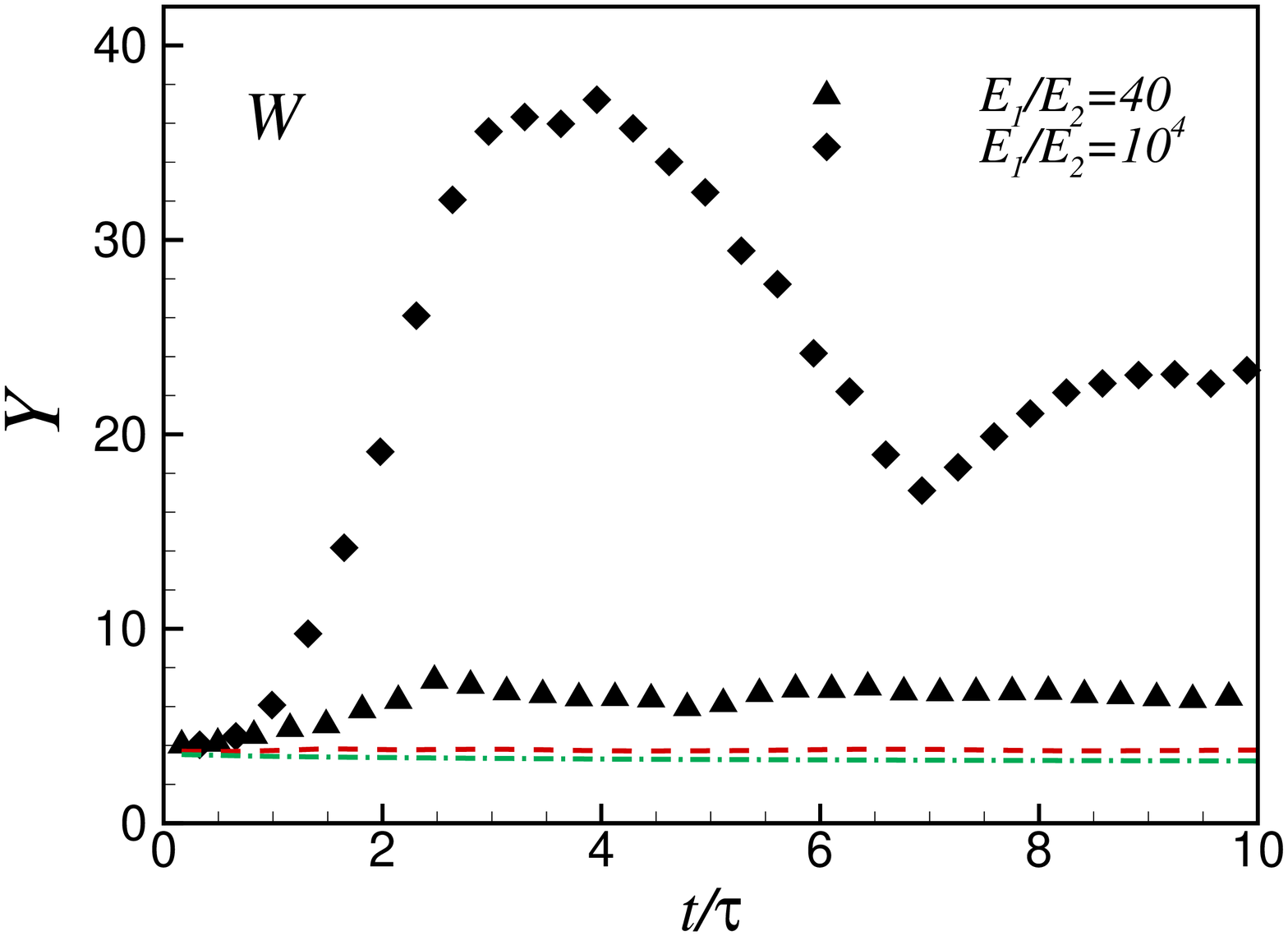}
\caption{(Color online) Temporal evolution of the peak of the velocity derivative statistics within the mixing.
(a) Longitudinal velocity dervative skewness.
(b) Longitudinal velocity derivative kurtosis. 
(c) Transverse velocity derivative skewness.
(d) Transverse velocity derivative kurtosis. The instantaneous values in the high and low homogeneous regions external to the mixing are shown with the red (dashed) and green (dash-dotted) lines respectively. The violet (dotted) line is the reference value for $Re_\lambda=45$ deduced from figures 5 and 6 in ref.\ \cite{sa97}.
}
\label{fig.derLT}
\end{figure}

An alternative measure of the anisotropy is given by the  velocity gradient statistics. We have computed the third and fourth order moments of both the longitudinal velocity derivative $\partial u/\partial x$ and transverse velocity derivative $\partial u/\partial {y_i}$ (no summation over $i$). These are so defined
\begin{eqnarray*}
S_{\partial u/\partial x}&=&\overline{(\partial u / \partial x)^3}/(\overline{(\partial u/\partial x)^2})^{3/2},\\
S_{\partial u/\partial {y_i}}&=&\overline{(\partial u/\partial y_i)^3}/(\overline{(\partial u/\partial {y_i})^2})^{3/2},
\;\;i=1,2
\end{eqnarray*}
\begin{eqnarray*}
K_{\partial u/\partial x}&=&\overline{(\partial u/\partial x)^4}/(\overline{(\partial u/\partial x)^2})^2,\\
K_{\partial u/\partial {y_i}}&=&\overline{(\partial u/\partial y_i)^4}/(\overline{(\partial u/\partial {y_i})^2})^2, 
\;\;i=1,2
\end{eqnarray*}
The averages are computed by integrating over planes at $x = const$. Figure \ref{fig.derLT} shows the time evolution of the peak of the longitudinal and transverse velocity derivative skewness and kurtosis within the mixing. The figure includes, for comparison, the values measured in the two homogeneous and isotropic turbulent fields outside the mixing and the values deduced from figures 5 and 6 of the review by Sreenivasan and Antonia (1997) \cite{sa97} for $Re_{\lambda}=45$.

We observe that the temporal evolution of all these velocity derivative statistics during the mixing decay presents an initial transient which is very similar to that shown by the velocity statistics, the transient length is the same in the two cases and there is no lag. The maximum values are always reached at $t/\tau \sim 4$ and increase with ${\cal E}$.  The longitudinal derivative moments are always larger than the transverse derivative moments, the difference decreases with the increase of ${\cal E}$. For instance, for ${\cal E} = 10^4$, absolute values as high as  $4 - 5$ are reached for the skewness, while values of $55$ and $38$  are measured for the longitudinal and transverse kurtosis, respectively. The anisotropy picture yielded by these velocity derivative correponds to that of a a higher intermittency along the inhomogeneous direction than across it.

\section{\label{conclusions}Conclusions}

We considered the simplest kind of turbulent shearless mixing process which is due to the interaction of two isotropic turbulent fields with different kinetic energy but the same spectrum shape. This mixing is characterized by  the absence of advection,   production of turbulent kinetic energy and an integral scale gradient. Such a situation can be seen as the simplest form of turbulence inhomogeneity that can lead to a departure from Gaussianity.  The  study was carried out by means of Navier-Stokes direct numerical simulations based on a fully dealiased Fou\-rier-Ga\-lerkin pseudospectral method of integration. The data base was analyzed through single-point statistics involving  the velocity and pressure fluctuations. 

   
We determined the temporal asymptotic behavior of the self-similar state. We also obtained the asymptotics for very high energy ratios between the isotropic turbulent fields which, through their interaction, initiate  the mixing process. The infinite limit of the turbulent energy ratio corresponds to the interaction of a region of isotropic turbulence with a relatively still fluid. In this limit the turbulent energy gradient reaches the  maximum observable value associated to a given energy in the high energy side of the mixing. In this limit the mixing penetration is maximum and is as deep as 1.2 times the mixing thickness. 

We observed the intermittency and anisotropy of the mixings.
Anisotropy was found to be  mild for second order moments,  on the contrary it was very intense in third and fourth order moments. The time asymptotic behavior of the anisotropy was almost independent of the turbulent energy ratio (i.e. turbulent energy gradient). The anisotropy observed through the third and fourth order moments of the velocity derivatives (longitudinal and transverse) is also very intense, but  depends on the turbulent energy ratio. 

Despite having no gradient of integral scale, no mean shear and thus no advection  and  no production of turbulent kinetic energy, all mixings showed a departure from a Gaussian state for any turbulent energy ratio. This signifies that the absence of these flow properties does not imply a condition of no intermittency. On the contrary the intermittency is highly dependant on the turbulent energy ratio between the two interacting fields. The intermittency has a constant asymptote when this ratio approaches to infinity, which is consistent with the maximum value of the turbulent energy gradient that can be asymptotically attained  in this limit. It is deduced that the presence of a gradient of turbulent kinetic energy is a sufficient condition for the onset of intermittency. For any turbulent energy ratio we verified that the pressure transport is not negligible with regard to the velocity transport as in recirculating turbulent flows.

In conclusion, by assuming that the interaction of two isotropic turbulent fields with different kinetic energy but the same  integral scale is the non-homogeneous turbulent flow with the lowest level of dynamical complexity, we propose  the hypothesis that the existence of a gradient of turbulent energy is the minimal requirement for Gaussian departure in turbulence, since there is experimental evidence that it is a sufficient condition to promote intermittency.

\begin{acknowledgments}
We wish to acknowledge the support of CINECA, HLRS and BCS supercomputing centers in supporting this work. 
This research project has been supported by the AeroTranet Marie Curie Early Stage
Research Training Fellowship of the European Community's Sixth Framework
Programme under contract number MEST CT 2005 020301.

\end{acknowledgments}

\appendix
\section{Mean pressure field in the turbulent shearless mixing flow}

The shearless turbulent mixing that we have studied is a flow where the average momentum is zero since the initial condition and the boundary conditions are such as to not generate a mean flow. It should be noted that the laboratory configuration, at least, those to date, is somehow different. In fact, when the two interacting homogeneous isotropic turbulent fields are generated by grids placed in a wind tunnel, a mean (homogeneous, i.e. shearless) flow is present in the normal direction to the mixing. However, it is true that an acceleration along the mixing direction could emerge if the initial gradients of mean pressure and  turbulent kinetic energy do not compensate. As in the laboratory situation, this mean flow would remain homogeneous, thus also in this case the mixing would be shearless (i.e. devoid of the production of turbulent kinetic energy).
Let us first  consider the averaged Navier-Stokes equations without the introduction of any model.
The mean momentum equation is:
\begin{equation}
\partial_t U_i + \partial_j U_iU_j = - (1/\rho) \partial_i P -\partial_j\overline{u_iu_j} + \nu \nabla^2 U_i
\label{eq.rey}
\end{equation}
where the capital letters denote mean quantities, the small letters fluctuations, and the overline denotes the statistical average.
For $t=0$ we have $U_i=0$ and the only non zero derivative is in the $x$ direction, so that these  equations reduce to
\begin{equation}
\partial_t U   = - (1/\rho) \partial_x P - \partial_x\overline{u^2},
\label{eq.rey2}
\end{equation}
\noindent where $U$ is the mean velocity in the mixing direction.
It can be seen that if the initial pressure gradient term balances the gradient of  the part of the initial turbulent kinetic energy associated to the fluctuations in the $x$ direction $(\overline{u^2}= 2/3 K)$, the acceleration term $\partial_t U$ is  zero. In such a situation, a mean field will be absent. On the contrary, for example in the hypothetical case of an initial  kinetic energy gradient facing a zero pressure gradient,  a mean homogeneous (without shear) flow will be generated.

\begin{figure}


\psfrag{x}{$x$ [m]}
\psfrag{uu}{$\rho\overline{u^2}$}
\psfrag{PP}{$P$}
\psfrag{P}{\hspace*{-10mm}$P-p_0$ [Pa], $\rho\overline{u^2}$ [Pa]}
\includegraphics[width=0.85\columnwidth]{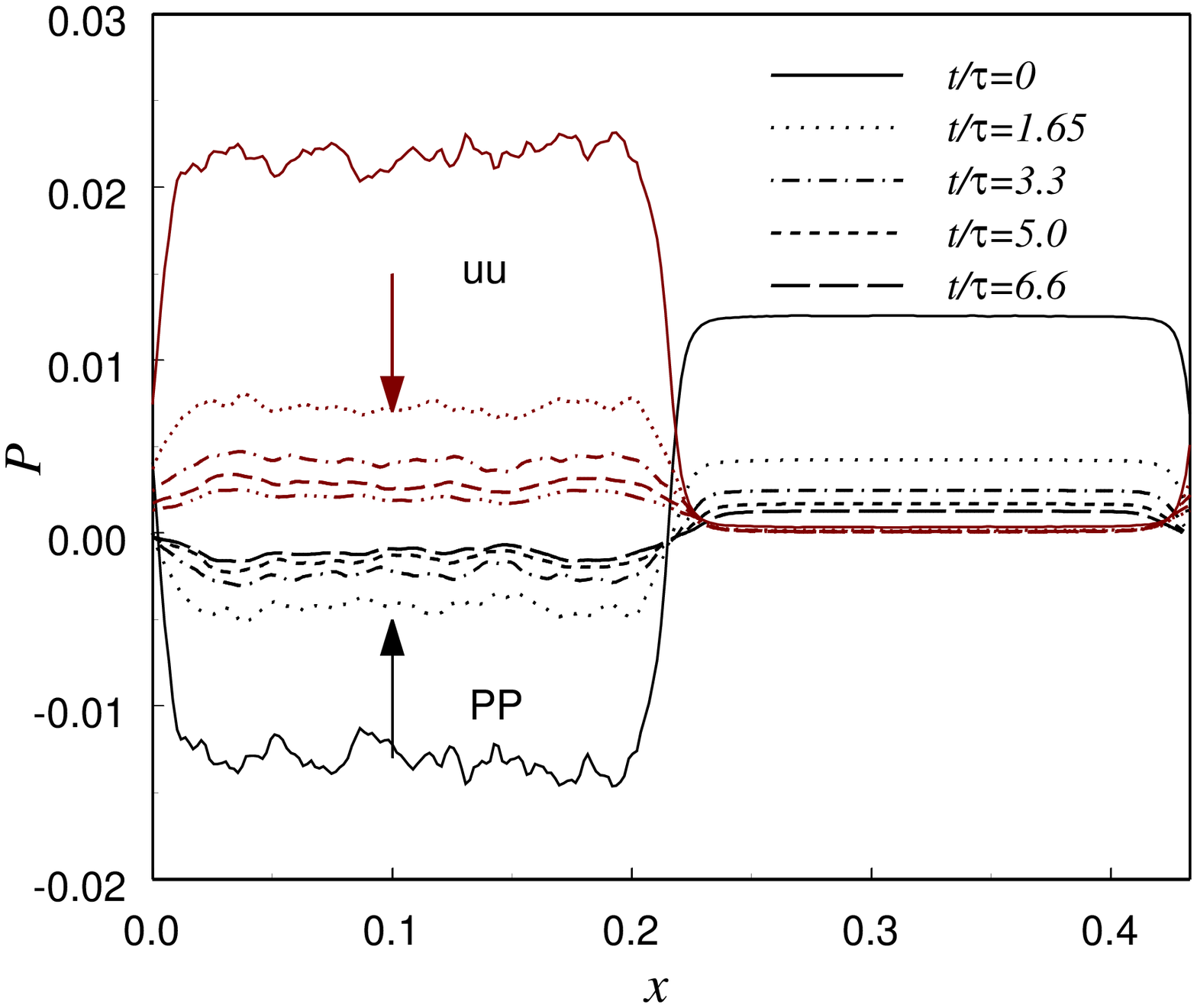}

\psfrag{dtu}{\footnotesize$\rho\partial_t U$ [Pa/m]}
\psfrag{duu}{\footnotesize$\rho \partial_x \overline{u^2}$ [Pa/m]}
\psfrag{dp}{\footnotesize$\partial_x P$ [Pa/m]}

\psfrag{t00}{$t/\tau=0$}
\psfrag{t33}{$\!t/\tau=3.3$}

\includegraphics[width=0.85\columnwidth]{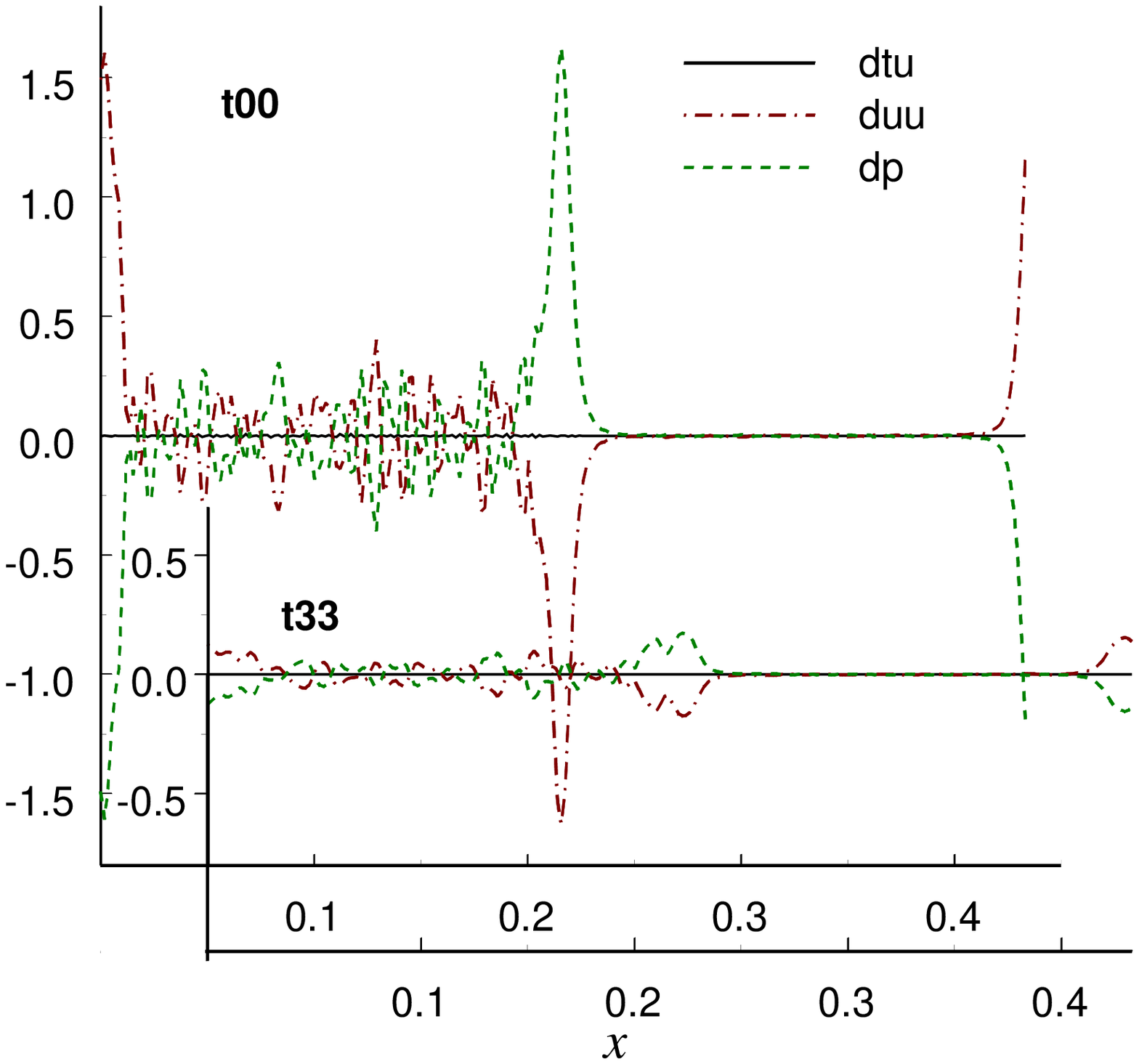}\\

\caption{(Color online) Part (a): profiles of the mean pressure and second order moment of the velocity component in the mixing direction, simulation with initial energy ratio $E_1/E_2=60$. Part (b): mean flow acceleration, gradient of the second order moment of the velocity component in the mixing direction and of the mean pressure. Simulation with initial energy ratio $E_1/E_2=60$. Air at standard conditions.} 
\label{fig.appendice}

\end{figure}

In the present numerical experiment the
initial velocity field is first introduced. Then, as is standard practice, the code builds the pressure field by using the Poisson equation obtained from the divergence of the momentum balance. Periodicity conditions plus a condition fixing the average pressure $p_0$ value in the entire domain are used.
Since the field is incompressible, the divergence of $U_i$ is zero and we obtain the following averaged equation:
\begin{equation}
\nabla^2 P /\rho = -\partial_i\partial_j\overline{u_iu_j} - \partial_i\partial_j U_iU_j
\end{equation}
%
At $t=0$ the fluctuating velocity field is statistically uniform apart from in the $x$ direction (note: it remains so during the mixing process). By also considering  the symmetries of the initial velocity field, and in particular the fact that, outside the mixing, the field is uniform, we obtain
\begin{equation}
\partial_{xx}^2 P /\rho = -\partial_{xx}^2\overline{u^2},  \;\; \partial_{x} P /\rho = -\partial_{x}\overline{u^2}.
\end{equation}
\noindent Consequently, by coming back  to (\ref{eq.rey2}), one can see that no mean acceleration is generated at $t=0$. 
Figure \ref{fig.appendice}  shows the terms in equation (A2) - the pressure and turbulent kinetic energy gradients and $\partial_t U$ - in two instants. 
We have considered the field configuration observed in the laboratory experiment by Veeravalli and Warhaft (1989, 3:1 perforated plate experiment, air flow at standard conditions), which is actually the field configuration that we tried to reproduce in this numerical experiment. In particular, we have estimated the dimensional values of the pressure gradients and pressure difference between the high turbulent energy and low energy regions of the mixing. 
\noindent If $({\rm d}P/{\rm d}x)_{max}=$ is the maximum value of the mean pressure gradient and $\Delta P=$ the pressure difference between the two homogeneous regions, we have at the initial instant of the simulations:\\[9pt]
%
%
$E_1/E_2=6.6, \;\; ({\rm d}P/{\rm d}x)_{max}=1.39 {\rm Pa/m}, \;\;$\\ $ \Delta P=2.30 \times 10^-2  \;\; {\rm Pa}, \;\; 2\Delta=2 {\rm cm}$\\[3pt]
$E_1/E_2=40, \;\;  \, ({\rm d}P/{\rm d}x)_{max}=1.60 {\rm Pa/m}, \;\; \,$\\ $ \Delta P=2.71 \times 10^-2 \;\; {\rm Pa}, \;\; 2\Delta=2 {\rm cm}$\\[3pt]
$E_1/E_2=60, \;\;  \, ({\rm d}P/{\rm d}x)_{max}=1.62 {\rm Pa/m}, \;\; \,$\\ $ \Delta P=2.72 \times 10^-2 \;\; {\rm Pa}, \;\; 2\Delta=2 {\rm cm}$\\[3pt]
%
%
\newline \noindent It can be observed that these pressure differences are very small. As a consequence,  measurements in the laboratory should  be very difficult.













\end{document}